\documentclass{article}
\date{Submitted for publication July 2007.\\ 
University of Colorado APPM preprint number 554.}
\usepackage{epsfig}
\usepackage{amsmath}
\usepackage{amssymb}
\usepackage{amsfonts}
\usepackage{array}

\newtheorem{thm}{Theorem} 

\newtheorem{defn}[thm]{Definition}
\newtheorem{rem}[thm]{Remark}

\newtheorem{prop}[thm]{Proposition}
\newcommand{\qed}{ \hfill$\square$}
\newenvironment{pf}{{\bf Proof:}}{\\}

\newcommand{\bigoh}{\mathcal{O} }
\newcommand{\vnoi}[1]{ \ensuremath{ {\bf  #1} } }
\newcommand{\mnoi}[1]{ \ensuremath{ {\mathbb  #1} } }
\newcommand{\op}[1]{ \ensuremath{ {\mathcal{#1}} } }

\newcommand{\srank}{ \ensuremath{{r}} }
\newcommand{\s}{ \ensuremath{s} }
\newcommand{\pv}{ \ensuremath{\vnoi{r}} }
\newcommand{\tv}{ \ensuremath{\gamma} }
\newcommand{\el}{ \ensuremath{N} }
\newcommand{\res}{{M}}
\newcommand{\ipmatrixwithi}{L}
\newcommand{\ipmatrix}{\mnoi{\ipmatrixwithi}}
\newcommand{\ipmatrixof}{{\ipmatrix}}
\newcommand{\gaussop}{\op{F}}
\newcommand{\wapply}[1]{{\op{W}_{_\op{P}}\left[ #1 \right]}}

\newcommand{\tvtimes}[1]{{(\op{T}_*+\op{V}_*)\left[ #1 \right]}}
\newcommand{\eval}[1]{{\delta({#1})}}
\newcommand{\aip}[1]{{\left\langle{#1}\right\rangle_{\op{A}}}}
\newcommand{\genop}{ \ensuremath{\op{L}} }
\newcommand{\orb}{\phi}
\newcommand{\aspi}[1]{ \ensuremath{ {#1} } }
\newcommand{\energy}{{E}}

\title{
Approximating a Wavefunction as an Unconstrained Sum of Slater Determinants%
}

\author{Gregory Beylkin\thanks{
Department of Applied Mathematics,
University of Colorado at Boulder,
526 UCB, Boulder CO 80309-0526;
\texttt{beylkin@colorado.edu}.
}
\and
Martin J. Mohlenkamp\thanks{
Department of Mathematics, 
Ohio University,
321 Morton Hall, Athens OH 45701;
\texttt{mjm@math.ohiou.edu}.
}
\and Fernando P\'erez\thanks{
Department of Applied Mathematics,
University of Colorado at Boulder,
526 UCB, Boulder CO 80309-0526;
\texttt{Fernando.Perez@colorado.edu}.
}
}
\begin{document} 

\maketitle
\markright{
an Unconstrained Sum of Slater Determinants 
\hfill}

\begin{abstract}
The wavefunction for the multiparticle Schr\"odinger equation is a
function of many variables and satisfies an antisymmetry condition, so
it is natural to approximate it as a sum of Slater determinants.  Many
current methods do so, but they impose additional structural
constraints on the determinants, such as orthogonality between
orbitals or an excitation pattern.  We present a method without any
such constraints, by which we hope to obtain much more efficient
expansions, and insight into the inherent structure of the
wavefunction. We use an integral formulation of the problem, a Green's
function iteration, and a fitting procedure based on the computational
paradigm of separated representations. The core procedure is the
construction and solution of a matrix-integral system derived from
antisymmetric inner products involving the potential operators. We
show how to construct and solve this system with computational
complexity competitive with current methods.
\end{abstract}


\clearpage
\section{Introduction}

Given the difficulties of solving the multiparticle Schr\"{o}dinger
equation, current numerical methods in quantum chemistry/physics are
remarkably successful. 
Part of their success comes from efficiencies gained by imposing
structural constraints on the wavefunction to match physical intuition.
However, such methods scale  poorly to high
accuracy, and are biased to only reveal structures that were part of
their own construction.  Our goal is to develop a method that
scales well to high accuracy and allows an unbiased exploration of the
structure of the wavefunction. 
In this paper we take a step toward this goal by developing a method
to approximate the wavefunction as an unconstrained sum of Slater
determinants. 

Since the multiparticle fermionic wavefunction is an antisymmetric
function of many variables, it is natural to approximate it as a sum
of Slater determinants, at least as a first step. Motivated by the
physical intuition that electrons may be excited into  higher energy
states, the Configuration Interaction (CI) family of methods choose
a set of determinants with predetermined orbitals, and then optimize
the coefficients used to combine them. When it is found
insufficient, methods to optimize the orbitals, work with multiple
reference states, etc., are introduced (along with an alphabet of
acronyms). A common feature of all these methods is that they impose
some structural constraints on the Slater determinants, such as
orthogonality of orbitals or an excitation pattern. As the requested
accuracy increases, these structural constraints trigger an explosion
in the number of determinants used, making the computation intractable
for high accuracy.

The a priori structural constraints present in CI-like methods also
force the wavefunction to comply with such structure, whether or not it
really is the case.  For example, if you use a method that
approximates the wavefunction as a linear combination of a reference
state and excited states, you could not learn that the wavefunction is
better approximated as a linear combination of several non-orthogonal,
near-reference states.  Thus the choice of numerical method is not
just a computational issue; it can help or hinder our understanding of
the wavefunction.

For these reasons, our goal is to  construct an adaptive numerical method
without imposing a priori structural constraints besides that of
antisymmetry.  In this paper we derive and present an algorithm for
approximating a wavefunction with an unconstrained sum of Slater
determinants, with fully-adaptive single-electron functions.
In particular we discard the notions of reference state
and excitation of orbitals.  The functions comprising the Slater
determinants need not come from a particular basis set, be orthogonal,
or follow some excitation pattern. They are computed so as to optimize
the overall representation. In this respect we follow the philosophy
of separated representations \cite{BEY-MOH:2002,BEY-MOH:2005}, which
allow surprisingly accurate expansions with remarkably few terms.

Our construction generates a solution using an iterative procedure
based on nonlinear approximations via separated representations.  To
accomplish this nonlinear approximation, we derive a system of
integral equations that describe the fully-correlated many-particle
problem. The computational core of the method is the repeated
construction and solution of a matrix-integral system of equations.

Specifically, our
approach has the following
distinctive features:
\begin{itemize}
\item
We use an adaptive representation for
single-electron functions, but our method does not depend on its details.
\item
We use an integral formulation of the multiparticle Schr\"{o}dinger
equation and a Green's function iteration to converge to the
ground-state wavefunction. The Green's function is decomposed and
applied using separated approximations obtained by expanding the
kernel into Gaussians.
\item
We use a variant of the so-called alternating least squares algorithm
to reduce the error of our approximation using a sum of a given number
of Slater determinants.
\item
We compute antisymmetric inner products involving portions of the
Hamiltonian operator by reducing them to formulas involving only
combinations of standard integrals. In particular, we avoid the direct
application of the electron-electron potential and instead compute
convolutions with the Poisson kernel.
\end{itemize}

By doing this, we hope to represent the effects of correlations in the
most natural and concise way possible, thus providing both
computational efficiency and physical insight.  We believe that this
algorithm and the system of integral equations underlying it provide
the foundation for a new approach to solving the multiparticle
Schr\"{o}dinger equation. We defer to the sequels several important
issues, such as algorithmic size-consistency/extensivity and the
treatment of the inter-electron cusp.

In Section~\ref{sec:description} we formulate the problem more
carefully, make precise some of the statements that we made in this
introduction, and give a high-level description of the method.
We then present the derivations and
proofs in the following sections.

\section{Problem Formulation and Description of the Method}
\label{sec:description}

\subsection{Formulation of the Problem}

We consider the time-independent, nonrelativistic, multiparticle
Schr\"odinger equation,   
and fix the nuclei according to the Born-Oppenheimer approximation, so
the equation describes the steady state of an interacting system of
electrons. For each of the $\el$ electrons in the system there are
three spatial variables $\pv=(x,y,z)$ and a discrete spin variable
$\sigma$ taking the values $\{-\frac{1}{2},\frac{1}{2}\}$, which we
combine and denote $(\pv,\sigma)$ by~$\tv$.  The Hamiltonian operator
$\op{H}$ is a sum of a kinetic energy operator $\op{T}$, a nuclear
potential operator $\op{V}$, and an electron-electron interaction
operator $\op{W}$, defined in atomic units by 
\begin{equation}
\op{H}=\op{T}+\op{V}+\op{W}=
-\frac{1}{2}  \sum_{i=1}^\el \Delta_i
+\sum_{i=1}^{\el} v(\pv_i)
+\frac{1}{2}\sum_{i=1}^{\el}\sum_{j\not= i}^{\el}
		\frac{1}{\|\pv_i-\pv_j\|}\,,
\end{equation}
where $\Delta_i$ is the three-dimensional Laplacian acting in the
variable $\pv_i$ and $v(\pv)$ is a sum of terms of the form
${-Z_a}/{\|\pv-\vnoi{R}_a\|}$ from a nucleus at position $\vnoi{R}_a$
with charge $Z_a$.  The antisymmetric eigenfunctions of $\op{H}$
represent electronic states of the system and are called
wavefunctions. Antisymmetric means that under the exchange of any two
coordinates, the wavefunction is odd, e.g.\
$\psi(\tv_2,\tv_1,\ldots)=-\psi(\tv_1,\tv_2,\ldots)$. 
The bound-state wavefunctions have negative eigenvalues, and are of
greatest interest.  We will focus on the ground-state wavefunction,
which has the most negative eigenvalue, although the techniques can be
used for other states.
In summary, our goal is to
find $\energy$ and $\psi$, with $\energy$ the most negative eigenvalue
in
\begin{equation}
\op{H}\psi=\energy\psi\,,
		\label{eqn:MPSeigen}
\end{equation}
subject to the antisymmetry condition on $\psi$.
Analytic methods can give qualitative
results about the solutions, and determine limiting cases, but most
quantitative results must be obtained numerically. 
Although the equation is `just' an eigenvalue problem,
its numerical solution presents several serious difficulties, among
them the large number of variables and the antisymmetry condition on
the solution. The simplest method that addresses these two difficulties
is Hartree-Fock (HF) (see e.g.\ \cite{HEL-TAY:1995a}), which uses the antisymmetrization of a
single product, called a {\em Slater determinant},  to approximate the $\el$-particle wavefunction, i.e.\
\begin{equation}
\psi_{\rm HF}= \op{A}\prod_{i=1}^{\el} \phi_i(\tv_i)
= \frac{1}{{\el!}}\left| \begin{array}{cccc}
\phi_1(\tv_1)	&\phi_1(\tv_2)	&\cdots	&\phi_1(\tv_\el)\\
\phi_2(\tv_1)	&\phi_2(\tv_2)	&\cdots	&\phi_2(\tv_\el)\\
\vdots		&\vdots		&	&\vdots\\
\phi_\el(\tv_1)	&\phi_\el(\tv_2)	&\cdots	&\phi_\el(\tv_\el)
	 \end{array}
\right|
\, .
		\label{eqn:HFpsi}
\end{equation}
Any antisymmetric approximation $\tilde{\psi}$ to the wavefunction $\psi$ can be substituted into
\begin{equation}
\frac{\langle \op{H} \tilde{\psi}, \tilde{\psi} \rangle}
	{\langle\tilde{\psi},\tilde{\psi}\rangle}
\,,
		\label{eqn:variationallambda}
\end{equation}
where $\langle \cdot,\cdot\rangle$ is the usual inner product,
to obtain an estimate for $\energy$. This estimate gives an
upper bound on the lowest value of $\energy$ that solves
(\ref{eqn:MPSeigen}). Substituting (\ref{eqn:HFpsi}) into
(\ref{eqn:variationallambda}), one can iteratively solve
for $\phi_i$ to minimize (\ref{eqn:variationallambda}). The resulting
$\psi_{\rm HF}$ will best approximate $\psi$, in the sense of
providing the best estimate (\ref{eqn:variationallambda}).

To improve upon HF, it is natural to consider the antisymmetrization
of a sum of products
\begin{equation}
\psi_{(\srank)}=\op{A}
\sum_{l=1}^\srank \s_l \prod_{i=1}^{\el} \phi_i^l(\tv_i)\, ,
			\label{eqn:psi_r}
\end{equation}
which could also be written as a sum of Slater determinants.  The
coefficients $\s_l$ are introduced in order to have
$\|\phi_i^l\|=1$. Many methods are based on this form, but they use it in
different ways. 
The Configuration Interaction (CI) method (see e.g.\
\cite{SHE-SCH:1999})
chooses the functions $\phi_i^l$ from a preselected master set of
orthogonal functions and decides on a large number $\srank$ of
combinations to consider, based on excitation level. Substituting
(\ref{eqn:psi_r}) into (\ref{eqn:variationallambda}) leads to a matrix
eigenvalue problem that can be solved for the scalar coefficients
$s_l$.
The Multi-Configuration Self-Consistent Field (MCSCF) method (e.g.\
\cite{GILBER:1972a,C-D-K-L-M:2003})
solves for the master
set of orthogonal functions as well as the scalar coefficients.
There are numerous variations and combinations of these methods, too
many to describe here. 

\subsubsection{What is New Here}

In this work we construct and demonstrate a method that also uses a
wavefunction of the form (\ref{eqn:psi_r}) but without constraints on
the $\phi_i^l$. We remove both structural constraints, such as an
excitation pattern or orthogonality between single-electron functions,
and representation constraints, such as those imposed by using a
predetermined basis set.

Many methods (e.g.\
\cite{RUDIN:1997,O-M-R-L-W:1987,LI-OL-RO:1988,AG-FL-JE:1986,GILBER:1972,DIJ-LEN:1998,FIN-STA:1993,AYA-SCH:1998,DIJ-LEN:2000,ZANGHE:2004,LUC-FIN:2000})
have loosened the constraints on the Slater determinants in one way or
another, often with encouraging results.  These works, however, only
partially removed the constraints, and so, we claim, did not achieve
the full potential of an unconstrained approximation.  By removing
these constraints we hope to produce much better approximations at
much smaller separation rank $\srank$ than existing methods allow. We
also hope to provide new perspective from which to analyze and
understand the wavefunction, free from the biases that physical
intuition imposes.

Our hopes are based on our work in 
\cite{BEY-MOH:2002,BEY-MOH:2005,MOH-MON:2005}, where we developed
general methods to represent and compute with functions and operators in many
dimensions. We used sums of separable functions, dubbed {\em separated
representations}, similar to (\ref{eqn:psi_r}). We found rather
natural  examples where removing constraints produces expansions
that are {\em exponentially}\/ more efficient, i.e.\ $\srank=\el$
instead of $2^{\el}$ or $\srank=\log \el$ instead of $\el$.
For example, in our
approach we can have a two-term representation
\begin{equation}
\psi=\op{A} \prod_{i=1}^\el\phi_i(\tv_i)
	+\op{A} \prod_{i=1}^\el(\phi_i(\tv_i)+\phi_{i+\el}(\tv_i))
\label{eqn:2vs2tod}
\end{equation}
where $\{\phi_{j}\}_{j=1}^{2\el}$ form an orthogonal set.  To
represent the same function as (\ref{eqn:2vs2tod}) while imposing the
constraint that factors come from a master orthogonal set would force
one to multiply out the second term, and thus use a representation
with $2^\el$ terms.

At present we have no proof that the wavefunction is well-approximated
by a structure that would benefit from the removal of constraints. The
size $\srank$ needed in practice, and how it depends on the various
parameters in the problem, is thus still an open question.
In
\cite{BEY-MOH:2002,BEY-MOH:2005,MOH-MON:2005}, the most interesting
examples came from ``reverse-engineering'' the numerical results to
obtain formulas and proofs.
We therefore expect that the tools we provide here will allow an
exploration of the wavefunction, perhaps revealing unexpected
structure, and a strategy for a proof.

\subsection{Description of the Algorithm}
\label{sec:maindescription}

The removal of constraints in (\ref{eqn:psi_r}), and, thus,  the basis sets,
coefficients, and other structure that went along with them, also
eliminates the conventional strategies for constructing
(\ref{eqn:psi_r}) to minimize (\ref{eqn:variationallambda}).
It leads one to consider how one would compute the ground-state wavefunction 
if its numerical representation were not an issue. We choose to use an
integral iteration, which we sketch in
Section~\ref{sec:greendescription}. In
Appendix~\ref{sec:descentdescription} we sketch an alternative
iteration based on gradient descent.

To use the form (\ref{eqn:psi_r}) we must choose some value of
$\srank$, which determines the quality of the approximation. In
Section~\ref{sec:greenproject} we show how to incorporate a nonlinear
fitting step within the integral iteration in order to maintain fixed
$\srank$. Accomplishing this fitting requires a significant amount of
machinery, which makes up the body of the paper. Eventually one would
want to adaptively determine $\srank$, but we do not address that
issue here.

\subsubsection{A Green's Function Iteration}
\label{sec:greendescription}

The eigenvalue equation (\ref{eqn:MPSeigen}) contains the differential
operator $\op{H}$, which has both the discrete negative eigenvalue(s) that
we are interested in and unbounded, continuous, positive spectrum.  In
\cite{KALOS:1962,KALOS:1963} this differential equation was
reformulated as an integral equation, producing an operator with only
discrete, bounded spectrum. Such integral formulations are in general
far superior to differential formulations, since, e.g.\ numerical
noise is suppressed rather than amplified.  An iteration based on the
integral formulation with Green's functions was introduced in
\cite{KALOS:1962,KALOS:1963} and used in e.g.\
\cite{CHE-AND:1995,H-F-Y-G-B:2004}. 
A rigorous analysis of this iteration is given in
\cite{MOH-YOU:2006P} based on classical theorems from
\cite{HUNZIK:1966,KATO:1951,REE-SIM:1975,REE-SIM:1978,RELLIC:1942}.
In this section we review this
iteration, and then modify it in Section~\ref{sec:greenproject} to preserve our
wavefunction representation (\ref{eqn:psi_r}).

Define the Green's function
\begin{equation}
\op{G}_\mu=(\op{T}-\mu \op{I})^{-1}\, ,
	\label{eqn:Gmudef}
\end{equation}
for $\mu<0$, and consider the  Lippmann-Schwinger integral equation
\begin{equation}
\lambda_\mu\psi_\mu
=-\op{G}_{\mu}[(\op{V}+\op{W})\psi_\mu]\,.
		\label{eqn:MPSintegral}
\end{equation}
The subscript $\mu$ on $\lambda_\mu$ and $\psi_\mu$ are to emphasize
the dependence of the eigenvalues and eigenfunctions on $\mu$. The
operator $\op{G}_{\mu}[(\op{V}+\op{W})]$ is compact, so
(\ref{eqn:MPSintegral}) has only discrete spectrum.  If $\mu=\energy$,
then there is an eigenvalue $\lambda_\mu=1$ and the corresponding
eigenfunction $\psi_\mu$ of (\ref{eqn:MPSintegral}) is the desired
ground-state eigenfunction of (\ref{eqn:MPSeigen}), as one can see by
rearranging (\ref{eqn:MPSintegral}) into (\ref{eqn:MPSeigen}).
We note that other eigenfunctions may be
obtained by deflation.  

When $\mu=\energy$, $\lambda_\mu=1$ is the largest eigenvalue, so a
simple iteration like the power method yields the desired ground-state
eigenfunction.  
The eigenvalues $\lambda_\mu$ depend analytically on
$\mu$, so when $\mu$ is sufficiently close to $\energy$ the power
method will still yield an eigenfunction of (\ref{eqn:MPSintegral})
with energy near the minimum of (\ref{eqn:variationallambda}). From
$\psi_\mu$ and $\lambda_\mu$ one can construct an update rule for
$\mu$, based for example on applying Newton's method to solve
$\lambda_\mu=1$.

The convergence rate of the power method to produce $\psi_\mu$ and
$\lambda_\mu$ is linear, and depends, as usual, on the gap between the
eigenvalues in (\ref{eqn:MPSintegral}). The convergence rate of
Newton's method to solve $\lambda_\mu=1$ is quadratic, so $\mu$ will
converge to $\energy$ quadratically, provided that $\lambda_\mu$ and
$\psi_\mu$ have been found at each step.
In the practical use of this approach, one does not wait for the power
method to converge at each step, but instead intertwines 
it with the update of $\mu$.  Beginning with an approximation to the
energy $\mu_{0}\approx\energy$ and an approximate wavefunction
$\psi_{0}$, one converts (\ref{eqn:MPSintegral}) to an iteration
\begin{equation}
\tilde{\psi}_{n}=-\op{G}_{\mu_{n}}[(\op{V}+\op{W})\psi_{n}]\,.\label{eqn:Greeniter}
\end{equation}
After each iteration one normalizes by setting
\begin{equation}
\psi_{n+1} = \tilde{\psi}_{n}/\|\tilde{\psi}_{n}\|  
	\label{eqn:Greeniternorm}
\,.
\end{equation}
Following the approach
of  \cite{H-F-Y-G-B:2004}, we can use the update rule   
\begin{equation}
\mu_{n+1}  = \mu_n - \langle (\op{V}+\op{W})\psi_{n},\psi_n-\tilde{\psi}_n \rangle /\|\tilde{\psi}_{n}\|^2  
	\label{eqn:GreenitermuALT} 
\,,
\end{equation}
which is equivalent to using Newton's method.

\subsubsection{Approximating with Fixed Separation Rank $\srank$}
\label{sec:greenproject}

We restrict the method to approximate wavefunctions of the form
(\ref{eqn:psi_r}), with $\srank$ fixed, by replacing the definition of
$\tilde{\psi}_n$ in (\ref{eqn:Greeniter}).  We define $\tilde{\psi}_n$
to be the function of the form (\ref{eqn:psi_r}) that minimizes the
(least-squares) error
\begin{equation}
\|\tilde{\psi}_{n} -(- \op{G}_{\mu_n} [(\op{V}+\op{W}) \psi_{n}])\|.
	\label{eqn:LSerror}
\end{equation}
Since using (\ref{eqn:LSerror}) instead of (\ref{eqn:Greeniter})
introduces an error, the update rule (\ref{eqn:GreenitermuALT}) may no
longer give quadratic convergence, and in any case is not expected to
converge to the true energy. One may choose to replace the update rule
(\ref{eqn:GreenitermuALT}) with the more robust but slower converging
rule
\begin{equation}
\mu_{n+1}  = \frac{\langle \op{H} \psi_{n+1}, \psi_{n+1} \rangle}
	{\|\psi_{n+1} \|^2}
	\label{eqn:Greenitermu} 
\,,		
\end{equation}
which is based on (\ref{eqn:variationallambda}).
Other rules may be possible as well.
At present we do not have enough numerical experience to decide which
rule to prefer.

The Green's function iteration itself does not enforce the
antisymmetry condition.
In order to assure convergence to an antisymmetric solution, we use
the pseudo-norm induced by the pseudo inner product $\langle \cdot ,
\cdot \rangle_{\op{A}}=\langle\op{A}(\cdot) , \op{A}(\cdot) \rangle$,
as we did in \cite{BEY-MOH:2005}.  

The least-squares problem (\ref{eqn:LSerror}) is non-linear, and so
very difficult in general.  To simplify notation in the description of
our method, we now suppress the index $n$ in (\ref{eqn:LSerror}) and
consider a single problem of that form.  We begin by setting
$\tilde{\psi}=\psi$, and then iteratively improve $\tilde{\psi}$ to
reduce (\ref{eqn:LSerror}). Although we can see several strategies for
improving $\tilde{\psi}$, for concreteness we will restrict our description 
to the strategy most similar to \cite{BEY-MOH:2005}.
To improve the approximation $\tilde{\psi}$ we loop through the variables
(electrons). The functions in variables other than the current
variable are fixed, and the functions in the current variable are
modified to minimize the overall error (\ref{eqn:LSerror}). 
The error (\ref{eqn:LSerror}) depends linearly on the functions in a
single variable, so the minimization becomes much easier.
This
general Alternating Least-Squares
(ALS) approach is well-known (see e.g.\
\cite{HARSHM:1970,KR-PI-LE:1980,LE-MO-SI:1993,BRO:1997,LA-MO-VA:2000c,SM-GE-BR:2004}).
Although to minimize (\ref{eqn:LSerror}) one may need to loop through
the variables multiple times, it appears to be more cost effective to
loop only once and then do the next Green's function iteration. 
We alternate through the
directions, but for ease of exposition we describe the $k=1$ case. So,
$\tilde{\phi}_k^{l}$ is fixed for $k>1$, and we will solve for the values
of $\tilde{\phi}_1^{l}$ for all $l$.

To refine in the current variable, we set up and solve a linear
least-squares problem.  The normal equations for a least-squares
problem are derived by taking a gradient with respect to the free
parameters and setting the result equal to zero. As long as the
approximating function
is linear and not degenerate in these parameters, the resulting
equations are linear and have a unique solution, which minimizes
the error with respect to these parameters. 
Usually these free
parameters are coefficients of the representation in some fixed basis.
For example,
to find the coefficients $\{c_i\}$ to minimize 
\begin{equation}
\left\|f-\sum_i c_i g_i\right\|^2
=\left\langle f-\sum_i c_i g_i, f-\sum_i c_i g_i\right\rangle
\,,
\end{equation}
construct the normal equations
\begin{equation}
\mnoi{A} \vnoi{x}=\vnoi{b}\, ,
		\label{eqn:normal}
\end{equation}
with
\begin{equation}
A(k,i)=\langle g_k, g_i \rangle
\quad {\rm and}\quad
b(k)=\langle g_k, f\rangle
\,,
\end{equation}
solve them, and set $c_i=x(i)$.  Instead of using coefficients in some
basis as our parameters, we take the parameters to be the point values
of our functions $\tilde{\phi}_1^{l}$, so that the gradient becomes a
variational derivative. Formally,
we consider a
basis of delta functions $\eval{\tv-\cdot}$ and let their
coefficients be our parameters.
We still obtain linear normal equations (\ref{eqn:normal}),
but now $\vnoi{b}$ and $\vnoi{x}$ are vectors of functions, and
$\mnoi{A}$ is a matrix of integral operators. Specifically,
$b(l)$ is a function of $\tv$, $x(l')$ is a function of
$\tv'$, and $A(l,l')$ is an integral operator mapping
functions of $\tv'$ to functions of $\tv$.
The kernels in $\mnoi{A}$ are formally defined by 
\begin{equation}
A(l,l')(\tv,\tv')
=
\tilde{s}_l\tilde{s}_{l'}
\left\langle 
\eval{\tv-\tv_1}
\prod_{i=2}^\el \tilde{\phi}_i^{l}(\tv_i)
,
\eval{\tv'-\tv_1}
\prod_{i=2}^\el \tilde{\phi}_i^{l'}(\tv_i)
\right\rangle_{\op{A}}
\,,
		\label{eqn:normalAdef}
\end{equation}
and the functions in $\vnoi{b}$ are defined by 
\begin{equation}
b(l)(\tv)
=\tilde{s}_l\sum_{m}^{\srank} \s_{m}  
\left\langle 
\eval{\tv-\tv_1}
\prod_{i=2}^\el \tilde{\phi}_i^{l}(\tv_i)
,
- \op{G}_{\mu}[\op{V}+\op{W}]
\prod_{i=1}^\el \phi_i^{m}(\tv_i)
\right\rangle_{\op{A}}
\,.
			\label{eqn:bdefplain}
\end{equation}
Once we solve (\ref{eqn:normal}), we set $\tilde{\phi}_1^{l}=x(l)$. To
enforce the normalization convention $\|\tilde{\phi}_1^l\|=1$ we can
divide $\tilde{\phi}_1^{l}$ by its norm and incorporate the norm into
$\tilde{s}_l$. 

To solve the matrix-integral system (\ref{eqn:normal}), we need an
iterative method for solving linear systems that uses only operations
compatible with integral operators, such as matrix-vector products,
vector scales and additions, and vector inner products.
Typically the matrix $\mnoi{A}$ in normal equations is positive-definite.
Our operator $\mnoi{A}$ is only
semidefinite due to the nullspace in the antisymmetric
pseudonorm. Fortunately, $\vnoi{b}$ was computed with the same
pseudonorm and has no component in the nullspace of $\mnoi{A}$, so we
can still use methods for positive-definite matrices.
Based on these considerations, we choose to use
the Conjugate Gradient iterative method (see e.g.\
\cite{GOL-LOA:1996}) to solve (\ref{eqn:normal}). 
One initializes with $\vnoi{r}=\vnoi{b}-\mnoi{A}\vnoi{x}$,
$\vnoi{v}=\vnoi{r}$, and $c=\langle\vnoi{r},\vnoi{r}\rangle$, and then
the core of the method is the sequence of assignments
$\vnoi{z}\leftarrow\mnoi{A}\vnoi{v}$, $t\leftarrow
c/\langle\vnoi{v},\vnoi{z}\rangle$, $\vnoi{x}\leftarrow
\vnoi{x}+t\vnoi{v}$, $\vnoi{r}\leftarrow \vnoi{r}-t\vnoi{z}$,
$d\leftarrow \langle\vnoi{r},\vnoi{r}\rangle$, $\vnoi{v}\leftarrow
\vnoi{r}+(d/c)\vnoi{v}$, and $c\leftarrow d$, applied iteratively. 

One advantage of using this iterative method with integral operators
is that our algorithm does not rely on any particular basis.  The
representation of $\vnoi{x}$ can naturally be adaptive in $\tv$, for
example refining near the nuclei as indicated by the refinement in
$\vnoi{b}$. We assume the availability of some adaptive, high-accuracy
representation for single-electron functions, such as the polynomial
multiwavelet representation demonstrated in
\cite{H-F-Y-B:2003,H-F-Y-G-B:2004}, which effectively 
eliminates the basis-set error. For the estimates of computational
cost, we use $\res$ to denote the cost to represent a function of
$\tv$, or integrate such a function. The antisymmetry constraint
requires $\el \le \res$, and in general we expect $\res$ to be much
larger than $\el$.

\subsubsection{Summary of the Remainder of the Paper}

The core of the paper is the development of the methods needed to
construct $\mnoi{A}$ in (\ref{eqn:normalAdef}) and $\vnoi{b}$ in
(\ref{eqn:bdefplain}).
First, in Section~\ref{sec:Aip}, we develop the machinery and
algorithms for computing antisymmetric inner products involving the
operators $\op{T}$, $\op{V}$, and $\op{W}$.
Our formulation uses low-rank perturbations of matrices,
thus avoiding cofactor expansions. We also avoid explicit construction
of $\op{W}$ by incorporating its effect via spatial convolutions with the
Poisson kernel in three dimensions.
Second, in
Section~\ref{sec:Aipeval}, we show how to compute antisymmetric inner
products involving these operators and the delta function
$\eval{\tv-\tv_1}$. Again the key is to use low-rank
perturbations of matrices. 

In Section~\ref{sec:greendetail} we assemble all our tools to
demonstrate how to perform our main algorithm, and in particular how
to construct $\mnoi{A}$ in (\ref{eqn:normalAdef}) and $\vnoi{b}$ in
(\ref{eqn:bdefplain}).
We also gather the computational cost for the whole method. The cost
depends on the number of electrons $\el$, the separation rank
$\srank$, the one-particle representation cost $\res$, the number of
Green's function iterations $I$ (see
Section~\ref{sec:greendescription}),
and the
number of conjugate gradient iterations $S$ (see
Section~\ref{sec:greenproject}). Although $S$ in theory could be as
many as the number of degrees of freedom $\srank\res$, we have a very
good starting point, and so expect only a very small constant number
to be needed.  We use $\res\log\res$ to denote the generic cost to
convolve a function of $\tv$ with the Poisson kernel $1/\|\pv\|$. A
Fourier-based Poisson solver on a uniform grid would achieve this
complexity; for adaptive methods such as we use it is very difficult
to state the cost (see \cite{BE-CH-PE:2006P,ETH-GRE:2001}).  We use
$L$ to denote the number of terms used to approximate the Green's
function to relative error $\epsilon$ with Gaussians,
and prove in Section~\ref{sec:greenrepresent} that
$L=\bigoh((\ln\epsilon)^2)$ independent of $\mu$ and $\el$.
The final computational cost is
then
\begin{equation}
\bigoh( I \srank^2 \el^2
[
L(\el+\res\log\res)
+S(\el+\res)
]
).
\end{equation}
For comparison, the cost to evaluate a single antisymmetric inner product via L\"owdin's
rules is $\bigoh(\el^2(\el+\res))$.  

\subsection{Further Considerations}

We have implemented the method developed here and tested it
sufficiently to verify the correctness of the algorithm as presented.
The numerical results are too preliminary to allow us to make any
particular claims at this point, however, so we will present them
separately. The linear algebra accelerations based on
Appendix~\ref{sec:pseudoupdate} have not yet been implemented.

We develop the method in terms of the total variable $\tv$ without
specifying the spin states. If a specific spin state is imposed on
our initial trial wavefunction $\psi_0$, the iteration will preserve this
state.

The representation (\ref{eqn:psi_r}) does not account for the
inter-electron cusp (see e.g.\
\cite{RY-CE-KO:1994,NO-KU-KL:1992,KLO-SAM:2002,PER-TAY:1996,PER-TAY:1997,KLOPPE:2000,LEBRIS:2003}),
and thus we cannot hope to achieve small error $\epsilon$ in the
wavefunction with small $\srank$.  As with Configuration Interaction
methods, we may still be able to achieve small error in the energy
difference of two systems, which is often the quantity of interest in
Chemistry. For the current work, we fix $\srank$ and adapt
$\phi_i^l(\tv_i)$ and $\s_l$ to minimize the error $\epsilon$, rather
than fixing $\epsilon$ and adaptively determining $\srank$.  We are
developing an extension to (\ref{eqn:psi_r}) that incorporates the
cusp, and hope to achieve small error $\epsilon$ through it.


Similarly, (\ref{eqn:psi_r}) is not size-consistent/extensive, and
thus is not suitable for large systems. 
We are also developing an extension to (\ref{eqn:psi_r}) suitable for
large systems, and hope to achieve linear scaling through it.

Although we have focused on the multiparticle Schr\"odinger equation,
the tools that we have developed are another step towards general-purpose,
automatically adaptive methods for solving high-dimensional problems.

\section{Antisymmetric Inner Products}
\label{sec:Aip}

In this section we develop methods for 
computing antisymmetric inner products involving $\op{W}$,
$\op{V}$, and $\op{T}$.
For this purpose,
after setting notation, we develop methods for
computing with low rank perturbations of matrices,  review the
antisymmetry constraint and define a notion of maximum
coincidence. With these tools we then derive the main formulas. 

\subsection{Notation}

We denote a column {\em vector}\/ with suppressed indices by
$\vnoi{F}$ and with explicit indices by $F(i)$. We denote its
conjugate transpose by $\vnoi{F}^*$. We use $\vnoi{e}_i$ to denote the
column vector that is one in coordinate $i$ and zero otherwise.  A
linear {\em operator} is written $\genop$.  We denote a {\em matrix}\/
with suppressed indices by $\ipmatrix$ and with explicit indices by
$L(i,j)$.  Recalling that $\pv=(x,y,z)\in\vnoi{R}^3$, we combine
spatial integration with summation over spins and define the integral
\begin{equation}
\int f(\tv) d\tv= \sum_{\sigma\in\{-1/2,1/2\}} \int f(\pv,\sigma) d\pv.
\end{equation}
We define the action of the single-electron kinetic and nuclear
potential operators by
\begin{equation}
\tvtimes{f}(\tv)=\left(-\frac{1}{2}\Delta+v(\pv)\right)f(\tv)
=\left(-\frac{1}{2}\Delta+v(\pv)\right)f(\pv,\sigma).
\end{equation}
In what follows we will reduce the action of the inter-electron
potential operator $\op{W}$ to convolutions with the Poisson kernel, so we
define
\begin{equation}
\wapply{f}(\pv)=\int\frac{1}{\|\pv-\pv'\|} f(\tv') d\tv'
=\sum_{\sigma'\in\{-1/2,1/2\}}\int\frac{1}{\|\pv-\pv'\|} f(\pv',\sigma') d\pv'
\,.
\label{eqn:wapplydef}
\end{equation}
We allow these operators to be applied componentwise to vectors and
matrices of functions. 

Next, we define $\aspi{\Phi}=\prod_{i=1}^{\el}\phi_i(\tv_i)$, so 
for example we
can write $\langle\aspi{\tilde{\Phi}},\aspi{\Phi}\rangle_\op{A}$ instead of 
$\aip{\prod_{i=1}^{\el}\tilde{\phi}_i(\tv_i),\prod_{i=1}^{\el}\phi_i(\tv_i)}$.
We also associate with the product  $\aspi{\Phi}$
a vector of $\el$ functions of a single variable,
\begin{equation}
\vnoi{\Phi}=\left[\begin{array}{c}\phi_1\\\phi_2\\\vdots\\\phi_\el\end{array}\right]
\,.
		\label{eqn:Phiasvector}
\end{equation}
We can then, for example, construct a new vector of functions
$\vnoi{\Theta}$ by applying a matrix to an old one, as in
$\vnoi{\Theta}=\ipmatrix^{-1}\tilde{\vnoi{\Phi}}$.  Although we do
linear algebra operations on these vectors, we note that
$\vnoi{\Phi}+\tilde{\vnoi{\Phi}}$ does not correspond to
$\aspi{\Phi}+\aspi{\tilde{\Phi}}$, so there is not a true vector-space
structure.
Our formulas contain fairly complicated expressions with such vectors,
such as $\int 
\vnoi{\Phi}^*\wapply{\vnoi{\Theta}\vnoi{\Phi}^*}\vnoi{\Theta}
d\tv$. To parse this expression, 
we note that $\vnoi{\Theta}$ is a column vector of functions and
$\vnoi{\Phi}^*$ is a row vector of functions, so
$\vnoi{\Theta}\vnoi{\Phi}^*$ is a matrix of functions. Then
$\wapply{\vnoi{\Theta}\vnoi{\Phi}^*}$ is still a matrix of functions,
but applying $\vnoi{\Phi}^*$ on the left and $\vnoi{\Theta}$ on the right yields a
single function, which is integrated in the implied variable $\tv$ to
yield a number.
When explicit specification of the variable involved is needed,
the notation $\vnoi{\Phi}(\tv)$ indicates that
the single variable $\tv$ is used in all the functions.

\subsection{Determinants of Low-Rank Perturbations of Matrices}
\label{sec:LRdeterminant}

Since the antisymmetric inner product involves determinants, we will
use some linear algebra relations for
them. Proposition~\ref{pro:pertrankdet} in this section is used
heavily, and is the key to avoiding rather unpleasant cofactor expansions.

\begin{prop}[Determinant via Schur Complement]
Let $\mnoi{A}$ be a nonsingular square matrix, $\mnoi{D}$ a square
matrix, and $\mnoi{B}$ and $\mnoi{C}$ matrices of appropriate size. Then
\begin{equation}
\left|
\begin{array}{cc}
\mnoi{A}	&\mnoi{B}\\
\mnoi{C}	&\mnoi{D}
\end{array}
\right|
=
|\mnoi{A}|
\left|\mnoi{D}-\mnoi{C}\mnoi{A}^{-1}\mnoi{B}\right|\, .
\label{eqn:schurdet}
\end{equation}
\label{pro:schurdet}
\end{prop}
\begin{pf} 
(see e.g. \cite{PRASOL:1994}) It is easy to verify directly that
\begin{equation}
\left[
\begin{array}{cc}
\mnoi{A}	&\mnoi{B}\\
\mnoi{C}	&\mnoi{D}
\end{array}
\right]
=
\left[
\begin{array}{cc}
\mnoi{I}	&0\\
\mnoi{C}\mnoi{A}^{-1}	&\mnoi{I}
\end{array}
\right]
\left[
\begin{array}{cc}
\mnoi{A}	&0\\
0	&\mnoi{D}-\mnoi{C}\mnoi{A}^{-1}\mnoi{B}
\end{array}
\right]
\left[
\begin{array}{cc}
\mnoi{I}	&\mnoi{A}^{-1}\mnoi{B}\\
0	&\mnoi{I}
\end{array}
\right].
\end{equation}
Since the determinants of the first and third matrices are equal to one, the
determinant of the middle matrix gives the desired result.
\qed\end{pf}

\begin{prop}[Determinant of a 
Perturbation of the
Identity]
Let $\{\vnoi{u}_q\}_{q=1}^{Q}$ and $\{\vnoi{v}_q\}_{q=1}^{Q}$ be two
sets of vectors of the same length, and $\vnoi{u}_q\vnoi{v}_q^*$ denote the outer product
of $\vnoi{u}_q$, and $\vnoi{v}_q$.
Then
\begin{equation}
\left| \mnoi{I}+\sum_{q=1}^Q \vnoi{u}_q\vnoi{v}_q^* \right|
= 
\left| 
\begin{array}{cccc}
	1+\vnoi{v}_1^*\vnoi{u}_1 &\vnoi{v}_1^*\vnoi{u}_2 
		&\cdots & \vnoi{v}_1^*\vnoi{u}_Q\\
	\vnoi{v}_2^*\vnoi{u}_1 &1+\vnoi{v}_2^*\vnoi{u}_2 
		&\cdots & \vnoi{v}_2^*\vnoi{u}_Q\\
	\vdots & \vdots &\ddots & \vdots\\
	\vnoi{v}_Q^*\vnoi{u}_1 &\vnoi{v}_Q^*\vnoi{u}_2 
		&\cdots &1+ \vnoi{v}_Q^*\vnoi{u}_Q
\end{array}
\right|.
\label{eqn:pertrankdet}
\end{equation}
\label{pro:pertrankdet}
\end{prop}
\begin{pf}
Let $\mnoi{U}$ be the matrix with the vectors $\{\vnoi{u}_q\}$ as its
columns, and $\mnoi{V}$ the matrix with the vectors $\{\vnoi{v}_q\}$ as its
columns. Note that $\mnoi{U}$ and $\mnoi{V}$ are of the same size. By Proposition~\ref{pro:schurdet} we have
\begin{equation}
\left|
\begin{array}{cc}
\mnoi{I}	&\mnoi{U}\\
-\mnoi{V}^*	&\mnoi{I}
\end{array}
\right|
=
\left|\mnoi{I}+\mnoi{V}^*\mnoi{U}\right|,
\end{equation}
which evaluates to the right side of
(\ref{eqn:pertrankdet}). Exchanging the roles of $\mnoi{A}$ and
$\mnoi{D}$ in Proposition~\ref{pro:schurdet} we have
\begin{equation}
\left|
\begin{array}{cc}
\mnoi{I}	&\mnoi{U}\\
-\mnoi{V}^*	&\mnoi{I}
\end{array}
\right|
=
\left|\mnoi{I}+\mnoi{U}\mnoi{V}^*\right|,
\end{equation}
which evaluates to the left side of
(\ref{eqn:pertrankdet}).
\qed\end{pf}
The $Q=1$ case is well-known (see e.g.\ \cite{PRASOL:1994}) but 
we have not found the general case in the literature.

\subsection{The Modified Pseudo-inverse}

The singular value decomposition (SVD) (e.g.\ \cite{GOL-LOA:1996}) 
of a $\el\times\el$ matrix is 
\begin{equation}
\mnoi{A} = \sum_{i=1}^{\el} s_i \vnoi{u}_i \vnoi{v}_i^*=\mnoi{U}\mnoi{S}\mnoi{V}^*
\, ,
		\label{eqn:svd} 
\end{equation}
where the matrices $\mnoi{U}$ and $\mnoi{V}$ are unitary and
the singular values $\{s_i\}$ are non-neganive and in descending order.
The left singular vectors $\{\vnoi{u}_i\}$ form an orthonormal
set, as do the right singular vectors $\{\vnoi{v}_i\}$.
The pseudo-inverse is defined as
\begin{equation}
\mnoi{A}^{\dag} = \sum_{i=1}^{\el-Q} s_i^{-1} \vnoi{v}_i \vnoi{u}_i^*\, ,
\end{equation}
where $Q$ is the dimension of the (numerical) nullspace. We also
define a projection matrix onto the nullspace
\begin{defn}
\begin{equation}
\mnoi{A}^\perp=\sum_{i=\el-Q+1}^{\el}\vnoi{v}_i \vnoi{u}_i^* 
\end{equation}
\end{defn}
and a modified pseudo-inverse  
\begin{defn}[Modified Pseudo-Inverse]
\begin{equation}
\mnoi{A}^{\ddag} = \mnoi{A}^{\dag} + \mnoi{A}^\perp\,.
\end{equation}
		\label{def:Addag}
\end{defn}
Note that $\mnoi{A}^\perp$ and thus $\mnoi{A}^\ddag$ are not uniquely
defined since the choice of basis for the nullspace is not unique. For
our purposes any consistent choice works.
The modified pseudo-inverse behaves much like the
pseudo-inverse, but always has a non-zero determinant,
\begin{equation}
|\mnoi{A}^\ddag|=\left(|\mnoi{U}||\mnoi{V}^*|\prod_{s_i\not=0}s_i\right)^{-1}
\not=0\,.  
\end{equation}

\subsection{The Antisymmetrizer and L\"owdin's Rule}
\label{sec:antilowdin}

Given a separable function, its antisymmetric projection can be found
by applying the {\em antisymmetrizer\/} $\op{A}$ (see e.g.\
\cite{PAUNCZ:1995}), also called the {\em skew-symmetrization}\/ or
{\em alternation}\/ (see e.g.\ \cite{MUIR:1930,PRASOL:1994}), resulting in a
Slater determinant. In the vector notation (\ref{eqn:Phiasvector}), we have
\begin{equation}
\op{A}\aspi{\Phi}
=\frac{1}{{{\el!}}}\left|\left[
\begin{array}{ccc}\vnoi{\Phi}(\tv_1)&\cdots&\vnoi{\Phi}(\tv_{\el})\end{array}
\right]\right|
=\frac{1}{{{\el!}}}
\left| 
\begin{array}{cccc}
\orb_1(\tv_1)	&\orb_1(\tv_2)	&\cdots	&\orb_1(\tv_\el)\\
\orb_2(\tv_1)	&\orb_2(\tv_2)	&\cdots	&\orb_2(\tv_\el)\\
\vdots		&\vdots		&\ddots	&\vdots\\
\orb_\el(\tv_1)	&\orb_\el(\tv_2)	&\cdots	&\orb_\el(\tv_\el)\\
\end{array}
\right|.
\label{eqn:slater}
\end{equation}
One cannot explicitly form a Slater determinant $\op{A}\aspi{\Phi}$
for large $\el$ since it would have $\el!$ terms.  However, one can
compute the antisymmetric pseudo inner product
\begin{equation}
\langle \aspi{\tilde{\Phi}},\aspi{\Phi} \rangle_{\op{A}}
\overset{\rm def}{=} 
\langle \op{A}\aspi{\tilde{\Phi}}, \op{A} {\aspi{\Phi}}\rangle 
= \langle \aspi{\tilde{\Phi}}, \op{A} {\aspi{\Phi}}\rangle 
= \langle \op{A}\aspi{\tilde{\Phi}}, {\aspi{\Phi}}\rangle,
		\label{eqn:aipdef}
\end{equation}
where the first equality is a definition and the others follow since
$\op{A}$ is an orthogonal projector. It is not a true inner product
because it has a nullspace. To compute (\ref{eqn:aipdef}), first 
construct the matrix 
$\ipmatrix$
with entries
\begin{equation}
\ipmatrixwithi(i,j)=
\langle \tilde{\phi}_i,\phi_{j} \rangle
			\label{eqn:ipmatrixdef}
\end{equation}
at cost $\bigoh(\el^2\res)$. 
Then use 
$\langle \aspi{\tilde{\Phi}},\aspi{\Phi} \rangle_{\op{A}}
=\langle \op{A} \aspi{\tilde{\Phi}}, \aspi{\Phi} \rangle$ 
and  move the integrals inside the determinant to obtain
\begin{equation}
\langle \aspi{\tilde{\Phi}},\aspi{\Phi}\rangle_\op{A}
=\frac{1}{{\el!}}|\ipmatrix|\,,
\label{eqn:lowdin}
\end{equation}
which is the so-called 
L\"owdin's rule (e.g.\
\cite{LOWDIN:1955,PAUNCZ:1995}). 
Since $\ipmatrix$ is an ordinary matrix, its determinant can be
computed with cost $\bigoh(\el^3)$ (or less).  The denominator
${\el!}$ need never be computed, since it will occur in every term in
our equations, and so cancels.

Our method for enforcing the antisymmetry constraint, as described in
\cite{BEY-MOH:2005}, is to use the pseudo-norm based on
the antisymmetric inner product $\langle\cdot,\cdot\rangle_\op{A}$ for
the least-squares fitting (\ref{eqn:LSerror}).

\subsection{Maximum Coincidence}
\label{sec:coincide}

Consider two products, $\aspi{\Phi}=\prod_{i=1}^{\el}\phi_i(\tv_i)$
and $\tilde{\aspi{\Phi}}=\prod_{i=1}^{\el}\tilde{\phi}_i(\tv_i)$,
stored in the vector notation of (\ref{eqn:Phiasvector}) as
$\vnoi{\Phi}$ and $\tilde{\vnoi{\Phi}}$.  To specify which functions
were used to compute $\ipmatrix$ in (\ref{eqn:ipmatrixdef}), we use
the notation $\ipmatrixof{(\tilde{\aspi{\Phi}},\aspi{\Phi})}$.  The
matrix of inner products
$\ipmatrix=\ipmatrixof(\tilde{\aspi{\Phi}},\aspi{\Phi})$ is in general
full.  Defining
\begin{equation}
\vnoi{\Theta}
=\ipmatrix^{-1}\tilde{\vnoi{\Phi}}
\,,
			\label{eqn:thetadef}
\end{equation}
we have 
\begin{multline}
\op{A}\aspi{\Theta}
=
\frac{1}{{{\el!}}}
\left|\left[
\begin{array}{ccc}
(\ipmatrix^{-1}\tilde{\vnoi{\Phi}})(\tv_1)&\cdots&(\ipmatrix^{-1}\tilde{\vnoi{\Phi}})(\tv_{\el})
\end{array}
\right]\right|\\
=|\ipmatrix^{-1}|
\frac{1}{{{\el!}}}
\left|\left[
\begin{array}{ccc}
\tilde{\vnoi{\Phi}}(\tv_1)&\cdots&\tilde{\vnoi{\Phi}}(\tv_{\el})
\end{array}
\right]\right|
=|\ipmatrix^{-1}|\op{A}\aspi{\tilde{\Phi}}
\,.
\end{multline}
Thus the antisymmetrizations of $\aspi{\tilde{\Phi}}$ and
$\aspi{\Theta}$ are the same up to a constant, and we can use
$\aspi{\Theta}$ instead of $\aspi{\tilde{\Phi}}$ in calculations. The
advantage of using $\aspi{\Theta}$ is that the resulting matrix of inner products
$\hat{\ipmatrix}=\ipmatrixof(\aspi{\Theta},\aspi{\Phi})=\mnoi{I}$; in other words, we
have the biorthogonality property $\langle \theta_i,\phi_{j}
\rangle=\delta_{ij}$.
To show this, write the matrix $\hat{\ipmatrix}$ as $\int
\vnoi{\Theta} \vnoi{\Phi}^*d\tv$, where the integration is
elementwise. Substituting for $\vnoi{\Theta}$, we have $\int
(\ipmatrix^{-1}\tilde{\vnoi{\Phi}}) \vnoi{\Phi}^*d\tv$. Since the
integration is elementwise it commutes with $\ipmatrix^{-1}$ and we
have $\ipmatrix^{-1}\int\tilde{\vnoi{\Phi}}
\vnoi{\Phi}^*d\tv=\ipmatrix^{-1} \ipmatrix=
\mnoi{I}$. 
The computational cost to construct $\vnoi{\Theta}$ is $\bigoh(\el^2(\el+\res))$. 

When the matrix $\ipmatrix$ in (\ref{eqn:ipmatrixdef}) is singular, we
define
$\vnoi{\Theta}=\ipmatrix^{\ddag}\tilde{\vnoi{\Phi}}$ using the
modified pseudo-inverse of Definition~\ref{def:Addag}. By the same
argument as before, we have $|\ipmatrix^{\ddag}|^{-1}\op{A}
\aspi{\Theta}=\op{A}\aspi{\tilde{\Phi}}$. The matrix $\int
\vnoi{\Theta} \vnoi{\Phi}^*d\tv$ evaluates  
to $\ipmatrix^{\ddag} \ipmatrix=\mnoi{I}-\sum_{i=\el-Q+1}^{\el}
\vnoi{v}_i\vnoi{v}_i^*$. For notational convenience in later sections,
we will re-index our singular values and vectors so that the first
$Q$ generate the nullspace, rather than the last~$Q$.

\begin{rem}
Within Configuration Interaction methods, the functions in
$\vnoi{\Phi}$ and $\tilde{\vnoi{\Phi}}$ are taken from a master
set of orthonormal functions, and $\vnoi{\Theta}$ is simply a signed
permutation of $\tilde{\vnoi{\Phi}}$ so that $\phi_j=\theta_j$ for as many
$j$ as possible. This is known as the `maximum coincidence' ordering.
The construction we use generalizes this notion.
\end{rem}

\subsection{Antisymmetric Inner Product with the Electron-Electron
Potential $\op{W}$ Present}
\label{sec:AipW}

In this section we derive formulas for computing antisymmetric inner
products that include the electron-electron interaction
potential. Although the derivation is somewhat messy, the resulting
formulas are rather clean, and we use them verbatim in the
computations. The main ideas are given in this section, and then
reused in later sections for other cases.

\begin{prop}
When $\ipmatrix$ from (\ref{eqn:ipmatrixdef}) is nonsingular,
\begin{equation}
\aip{\aspi{\tilde{\Phi}},\op{W}\aspi{\Phi}}
\overset{\rm def}{=}
\left\langle
\op{A}\prod_{j=1}^\el \tilde{\phi}_j(\tv_j),
\left( \frac{1}{2}\sum_{i\not=j} \frac{1}{\|\pv_i-\pv_j\|} \right)\prod_{j=1}^\el \phi_j(\tv_j)
\right\rangle
		\label{eqn:AipW}
\end{equation}
is equal to
\begin{equation}
\frac{1}{2}\frac{|\ipmatrix|}{{\el!}}
\int
\vnoi{\Phi}^*\vnoi{\Theta}
\wapply{\vnoi{\Phi}^*\vnoi{\Theta}}
-
\vnoi{\Phi}^*\wapply{\vnoi{\Theta}\vnoi{\Phi}^*}\vnoi{\Theta}
d\tv 
\,,
		\label{eqn:AipWfinal}
\end{equation}
where $\vnoi{\Theta}=\ipmatrix^{-1}\tilde{\vnoi{\Phi}}$.
\end{prop}
\begin{pf}
Using the maximum-coincidence procedure in
Section~\ref{sec:coincide}, (\ref{eqn:AipW}) is equal to
$|\ipmatrix|\aip{\aspi{\Theta},\op{W}\aspi{\Phi}}$. 
We  reorganize and find that we must compute
\begin{equation}
\frac{1}{2}\frac{|\ipmatrix|}{{\el!}}
\int
\left( \sum_{i\not=j} \frac{1}{\|\pv_i-\pv_j\|} \right)
\prod_{j=1}^\el \overline{\phi}_j(\tv_j)
\left| 
\begin{array}{cccc}
\theta_1(\tv_1)	&\theta_1(\tv_2)	&\cdots	&\theta_1(\tv_\el)\\
\theta_2(\tv_1)	&\theta_2(\tv_2)	&\cdots	&\theta_2(\tv_\el)\\
\vdots		&\vdots		&\ddots	&\vdots\\
\theta_\el(\tv_1)	&\theta_\el(\tv_2)	&\cdots	&\theta_\el(\tv_\el)\\
\end{array}
\right|
d\tv_1\cdots d\tv_\el \, .
				\label{eqn:gAintegral}
\end{equation}
By moving the sum outside of the integral, we can
integrate in all directions except $\tv_i$ and $\tv_j$. Using $\langle
\theta_m,\phi_n\rangle=\delta_{mn}$, we obtain
\begin{multline}
\frac{1}{2}\frac{|\ipmatrix|}{{\el!}}
\sum_{i\not=j}
\int
\frac{1}{\|\pv-\pv'\|}
\left| 
\begin{array}{ccccccc}
1	
	&\cdots	
	&\overline{\phi}_{i}(\tv)\theta_1(\tv)
	&\cdots
	&\overline{\phi}_{j}(\tv')\theta_1(\tv')
	&\cdots
	&0\\
\vdots
	&\ddots
	&\vdots
	&
	&\vdots
	&
	&\vdots\\
0	
	&\cdots	
	&\overline{\phi}_{i}(\tv)\theta_i(\tv)
	&\cdots
	&\overline{\phi}_{j}(\tv')\theta_i(\tv')
	&\cdots
	&0\\
\vdots
	&
	&\vdots
	&\ddots
	&\vdots
	&
	&\vdots\\
0	
	&\cdots	
	&\overline{\phi}_{i}(\tv)\theta_j(\tv)
	&\cdots
	&\overline{\phi}_{j}(\tv')\theta_j(\tv')
	&\cdots
	&0\\
\vdots
	&
	&\vdots
	&
	&\vdots
	&\ddots
	&\vdots\\
0	
	&\cdots	
	&\overline{\phi}_{i}(\tv)\theta_\el(\tv)
	&\cdots
	&\overline{\phi}_{j}(\tv')\theta_\el(\tv')
	&\cdots
	&1
\end{array}
\right|
d\tv d\tv' 
\\
=
\frac{1}{2}\frac{|\ipmatrix|}{{\el!}}
\sum_{i\not=j}
\int
\frac{1}{\|\pv-\pv'\|}
\bigg| \mnoi{I}
+\left(\overline{\phi}_{i}(\tv)
\vnoi{\Theta}(\tv)
-\vnoi{e}_i\right)
\vnoi{e}_i^*
+\left(\overline{\phi}_{j}(\tv')
\vnoi{\Theta}(\tv')
-\vnoi{e}_j\right)
\vnoi{e}_j^*
\bigg|
d\tv d\tv' \,.
			\label{eqn:Aip2I}
\end{multline}
Since the inner matrix is a low-rank perturbation of the identity, we
reduce its determinant using Proposition~\ref{pro:pertrankdet} and
obtain
\begin{equation}
\frac{1}{2}\frac{|\ipmatrix|}{{\el!}}
\sum_{i\not=j}
\int
\frac{1}{\|\pv-\pv'\|}
\overline{\phi}_{i}(\tv)\overline{\phi}_{j}(\tv')
\left| 
\begin{array}{cc}
	\theta_i(\tv)& \theta_i(\tv')\\
	\theta_j(\tv)& \theta_j(\tv')
\end{array}
\right|
d\tv d\tv' \,.
		\label{eqn:Aip1212}
\end{equation}
The determinant is zero if $j=i$, so we do not need to
explicitly prohibit it as we needed to in 
(\ref{eqn:Aip2I}) and above. The antisymmetrization has caused a convenient
cancellation of a fictitious self-interaction, and, thus, allowed us to
decouple the two sums.
Expanding out the determinant and rearranging the terms, we obtain 
\begin{multline}
\frac{1}{2}\frac{|\ipmatrix|}{{\el!}}
\int
\left(\sum_i 
\overline{\phi}_{i}(\tv)\theta_i(\tv)\right)
\left[
\int
\frac{1}{\|\pv-\pv'\|}
\left(\sum_j\overline{\phi}_{j}(\tv')\theta_j(\tv')\right)
d\tv' \right]
d\tv 
\\
-
\frac{1}{2}\frac{|\ipmatrix|}{{\el!}}
\int
\sum_i \sum_j
\overline{\phi}_{i}(\tv)
\theta_j(\tv)
\left[
\int
\frac{1}{\|\pv-\pv'\|}
\overline{\phi}_{j}(\tv')
\theta_i(\tv')
d\tv' \right]
d\tv 
\,.
\end{multline}
In our compact notation, this yields (\ref{eqn:AipWfinal}).
\qed\end{pf}

We now consider the computational cost of (\ref{eqn:AipWfinal}). 
In the first term in (\ref{eqn:AipWfinal}), computing
$\vnoi{\Phi}^*\vnoi{\Theta}$ costs $\bigoh(\el\res)$, applying
$\wapply{\cdot}$ to it costs $\bigoh(\res\log\res)$, and the integral in
$\tv$ costs $\bigoh(\res)$.  In the second term,
$\vnoi{\Phi}\vnoi{\Theta}^*$ costs $\bigoh(\el^2\res)$, applying
$\wapply{\cdot}$ to it costs $\bigoh(\el^2\res\log\res)$, applying
$\vnoi{\Theta}^*$ and then $\vnoi{\Phi}$ costs $\bigoh(\el^2\res)$,
and then the integral in $\tv$ costs $\bigoh(\res)$.  Including the
cost to construct $\vnoi{\Theta}$, our total cost is
$\bigoh(\el^2(\el+\res\log\res))$.

\subsubsection{The Singular Case}
\label{sec:AipWsingular}

In this section we investigate the case when the matrix $\ipmatrix$
from (\ref{eqn:ipmatrixdef}) is singular. Inserting the definition
$\vnoi{\Theta}=\ipmatrix^{-1}\tilde{\vnoi{\Phi}}$ into our main formula
(\ref{eqn:AipWfinal}), we have
\begin{equation}
\frac{1}{2}\frac{|\ipmatrix|}{{\el!}}
\int
\vnoi{\Phi}^*\ipmatrix^{-1}\tilde{\vnoi{\Phi}}
\wapply{\vnoi{\Phi}^*\ipmatrix^{-1}\tilde{\vnoi{\Phi}}
}
-
\vnoi{\Phi}^*\wapply{\ipmatrix^{-1}\tilde{\vnoi{\Phi}}\vnoi{\Phi}^*}\ipmatrix^{-1}\tilde{\vnoi{\Phi}}
d\tv 
\,.
		\label{eqn:AipWfinalLexplicit}
\end{equation}
In terms of the SVD (\ref{eqn:svd}), we can express
\begin{equation}
\ipmatrix^{-1}=\sum_{j=1}^{\el} s_j^{-1} \vnoi{v}_j \vnoi{u}_j^*
\quad {\rm and}\quad
|\ipmatrix|=|\mnoi{U}||\mnoi{V}^*|\prod_i s_i
\,.
\end{equation}
Inserting these expressions into (\ref{eqn:AipWfinalLexplicit}), we have
\begin{multline}
\frac{1}{2}\frac{|\mnoi{U}||\mnoi{V}^*|\prod_i s_i}{{\el!}}
\int
\vnoi{\Phi}^*\sum_{j=1}^{\el} s_j^{-1} \vnoi{v}_j \vnoi{u}_j^*\tilde{\vnoi{\Phi}}
\wapply{\vnoi{\Phi}^*\sum_{k=1}^{\el} s_k^{-1} \vnoi{v}_k \vnoi{u}_k^*\tilde{\vnoi{\Phi}}}\\
-
\vnoi{\Phi}^*
\wapply{\sum_{j=1}^{\el} s_j^{-1} \vnoi{v}_j \vnoi{u}_j^*\tilde{\vnoi{\Phi}}\vnoi{\Phi}^*}
\sum_{k=1}^{\el} s_k^{-1} \vnoi{v}_k \vnoi{u}_k^*\tilde{\vnoi{\Phi}}
d\tv \\
=
\frac{1}{2}\frac{|\mnoi{U}||\mnoi{V}^*|}{{\el!}}
\sum_{j=1}^{\el} \sum_{k=1}^{\el} 
\prod_{i\not=j,k} s_i
\int
\vnoi{\Phi}^*\vnoi{v}_j \vnoi{u}_j^*\tilde{\vnoi{\Phi}}
\wapply{\vnoi{\Phi}^* \vnoi{v}_k \vnoi{u}_k^*\tilde{\vnoi{\Phi}}}
\\
-
\vnoi{\Phi}^*\vnoi{v}_j 
\wapply{\vnoi{u}_j^*\tilde{\vnoi{\Phi}}
\vnoi{\Phi}^*  \vnoi{v}_k
}\vnoi{u}_k^*\tilde{\vnoi{\Phi}}
d\tv 
\,.
		\label{eqn:AipWasSVD}
\end{multline}
If $\ipmatrix$ is singular then at least one $s_i$ is zero, and only
terms that exclude those from the product in (\ref{eqn:AipWasSVD}) are
nonzero. Since we exclude
two indices in the product, if more than two $s_i$ are zero then the
entire inner product is zero. If exactly two are zero then only one
term in the sum survives. If exactly one is zero then we can simplify
from a double to a single sum, using symmetry. 
Recalling the modified pseudo inverse from
Definition~\ref{def:Addag} and sorting the zero $s_i$ to the beginning for notational
convenience, we obtain the following propositions.
\begin{prop}
When the rank-deficiency of $\ipmatrix$ is more than two, the
antisymmetric inner product
(\ref{eqn:AipW}) evaluates to zero.
\end{prop}
\begin{prop}
When the rank-deficiency  of $\ipmatrix$ is equal to  two,  the
antisymmetric inner product 
(\ref{eqn:AipW}) is equal to
\begin{equation}
\frac{1}{|\ipmatrix^{\ddag}|{\el!}}
\int 
\vnoi{\Phi}^*\vnoi{v}_1
\vnoi{u}_1^*\tilde{\vnoi{\Phi}}
\wapply{\vnoi{\Phi}^*\vnoi{v}_2\vnoi{u}_2^*\tilde{\vnoi{\Phi}}}
-
\vnoi{\Phi}^*\vnoi{v}_1
\wapply{\vnoi{\Phi}^*\vnoi{v}_2\vnoi{u}_1^*\tilde{\vnoi{\Phi}}}
\vnoi{u}_2^*\tilde{\vnoi{\Phi}}
d\tv 
\,.
			\label{eqn:AipWQ2final}
\end{equation}
\label{pro:AipWQ2}
\end{prop}
\begin{prop}
When the rank-deficiency of $\ipmatrix$ is equal to one, defining
$\vnoi{\Theta}=\ipmatrix^\dag\tilde{\vnoi{\Phi}}$ or
$\vnoi{\Theta}=\ipmatrix^\ddag\tilde{\vnoi{\Phi}}$,  the
antisymmetric inner product (\ref{eqn:AipW})
is equal to
\begin{equation}
\frac{1}{|\ipmatrix^{\ddag}|{\el!}}
\int
\vnoi{\Phi}^*\vnoi{v}_1
\vnoi{u}_1^*\tilde{\vnoi{\Phi}}
\wapply{\vnoi{\Phi}^*\vnoi{\Theta}}
-
\vnoi{\Phi}^*\vnoi{v}_1
\wapply{\vnoi{u}_1^*\tilde{\vnoi{\Phi}}\vnoi{\Phi}^*}
\vnoi{\Theta}
d\tv
\, .
			\label{eqn:AipWQ1final}
\end{equation}
\end{prop}

In computing (\ref{eqn:AipWQ2final}),
constructing $\vnoi{\Phi}^*\vnoi{v}_1$, $\vnoi{\Phi}^*\vnoi{v}_2$,
$\vnoi{u}_1^*\tilde{\vnoi{\Phi}}$, and
$\vnoi{u}_2^*\tilde{\vnoi{\Phi}}$ costs $\bigoh(\el\res)$, applying
$\wapply{\cdot}$ costs $\bigoh(\res\log\res)$ and, finally, the integral in
$\tv$ costs $\bigoh(\res)$. 
In computing (\ref{eqn:AipWQ1final}), 
the first term  costs $\bigoh(\el\res)$ to
form $\vnoi{\Phi}^*\vnoi{\Theta}$, $\bigoh(\res\log\res)$ to apply
$\wapply{\cdot}$, and $\bigoh(\res)$ to integrate in $\tv$. The second
term costs $\bigoh(\el\res)$ to form
$\vnoi{u}_1^*\tilde{\vnoi{\Phi}}\vnoi{\Phi}$,
$\bigoh(\el\res\log\res)$ to apply $\wapply{\cdot}$, $\bigoh(\el\res)$
to apply $\vnoi{\Theta}$, and $\bigoh(\res)$ to integrate in $\tv$.
In total, the computational cost for the singular cases are less than
the cost of the nonsingular case.

\begin{rem}
In the Configuration Interaction  context, rank-deficiency two
corresponds to a double excitation. The 
vectors $\vnoi{u}_i$ and $\vnoi{v}_i$ would be zero except for a single
entry, and so select the locations of the excited electrons out of
$\vnoi{\Phi}$ and  $\tilde{\vnoi{\Phi}}$. Proposition~\ref{pro:AipWQ2} then reduces to
the Slater-Condon rules \cite{CON-SHO:1935}.
\end{rem}

\subsection{Antisymmetric Inner Product with $\op{T}$ and/or $\op{V}$ Present}
\label{sec:AipV}
\label{sec:AipT}

Since $\op{T}$ and $\op{V}$ both have the structure of 
a sum of one-directional operators, we state the formulas
for their sum, although of course they can be treated individually.
\begin{prop}
If $\ipmatrix$ from (\ref{eqn:ipmatrixdef}) is nonsingular,
\begin{equation}
\aip{\aspi{\tilde{\Phi}},(\op{T}+\op{V})\aspi{\Phi}}
\overset{\rm def}{=}
\left\langle
\op{A}\prod_{j=1}^\el \tilde{\phi}_j(\tv_j),
\left( \sum_{i}-\frac{1}{2}\Delta_i+  v(\pv_i) \right)\prod_{j=1}^\el \phi_j(\tv_j)
\right\rangle
		\label{eqn:AipTV}
\end{equation}
is equal to
\begin{equation}
\frac{|\ipmatrix|}{{\el!}}
\int \tvtimes{\vnoi{\Phi}}^*\vnoi{\Theta} d\tv
 \,.
		\label{eqn:AipTVresult}
\end{equation}
\end{prop}
\begin{pf}
We follow the same procedure as we used for the electron-electron
operator $\op{W}$ in
Section~\ref{sec:AipW}. Instead of (\ref{eqn:Aip2I}) we have the
simpler expression
%
%
\begin{equation}
\frac{|\ipmatrix|}{{\el!}}
\sum_{i}
\int
\bigg| \mnoi{I}
+\left(\tvtimes{\overline{\phi}_{i}}(\tv)
\vnoi{\Theta}(\tv)
-\vnoi{e}_i\right)
\vnoi{e}_i^*
\bigg|
d\tv \,.
			\label{eqn:Aip1I}
\end{equation}
%
Applying Proposition~\ref{pro:pertrankdet}   
we obtain (\ref{eqn:AipTVresult}).
\qed\end{pf}

To analyze the computational cost to compute (\ref{eqn:AipTVresult}),
we note that 
it costs $\bigoh(\el\res)$ to apply $\tvtimes{\cdot}$. Including the
cost for the maximum coincidence transformation, our total cost is
thus $\bigoh(\el^2(\el+\res))$.

\subsubsection{The Singular Case}

We now state the formula when $\ipmatrix$ is singular.  
The analysis is similar to that for $\op{W}$ in
Section~\ref{sec:AipWsingular}. 

\begin{prop}
If the rank-deficiency of $\ipmatrix$ is greater than one, 
(\ref{eqn:AipTV}) evaluates to zero. 
If it is equal to one we have
\begin{equation}
\frac{1}{|\ipmatrix^{\ddag}|{\el!}}
\int \tvtimes{\vnoi{\Phi}^*\vnoi{v}_1} \vnoi{u}_1^*\tilde{\vnoi{\Phi}} d\tv
\, .
		\label{eqn:AipVQ1final}
\end{equation}
\end{prop}
To compute (\ref{eqn:AipVQ1final}),
it costs $\bigoh(\el\res)$ to form
$\vnoi{\Phi}^*\vnoi{v}_1$ and $\vnoi{u}_1^*\tilde{\vnoi{\Phi}}$, and
$\bigoh(\res)$ to apply $\tvtimes{\cdot}$.


\section{Incorporating Delta Functions into the Antisymmetric Inner Products}
\label{sec:Aipeval}

In this section we show how to compute antisymmetric inner products
when one of the component functions is replaced by a delta
function. For concreteness, we will replace $\tilde{\phi}_1(\tv_1)$ by
$\eval{\tv-\tv_1}$.

\subsection{L\"owdin's Rule with $\eval{\tv-\tv_1}$ Present}

The matrix $\ipmatrix$ from (\ref{eqn:ipmatrixdef}) is defined by
$\ipmatrixwithi(i,j)=\langle \tilde{\phi}_i,\phi_{j} \rangle$.  If we replace 
$\tilde{\phi}_1(\tv_1)$ by
$\eval{\tv-\tv_1}$, then the first row depends on $\tv$ and is given by
$\ipmatrixwithi(1,j)=\langle \eval{\tv-\cdot},\phi_{j}
\rangle=\phi_{j}(\tv)$. We thus have a matrix that depends on $\tv$, 
\begin{equation} 
\ipmatrix(\tv)=	
	\left[ 
\begin{array}{cccc}
\phi_1(\tv)	&\phi_2(\tv)	&\cdots	&\phi_\el(\tv)\\
\langle \tilde{\phi}_2,\phi_1\rangle
	&\langle \tilde{\phi}_2,\phi_2\rangle
	& \cdots
	&\langle \tilde{\phi}_2,\phi_{\el}\rangle\\
\vdots	&\vdots 	&\ddots			&\vdots\\
\langle \tilde{\phi}_{\el},\phi_1\rangle
	&\langle \tilde{\phi}_{\el},\phi_2\rangle
	& \cdots
	&\langle \tilde{\phi}_{\el},\phi_{\el}\rangle
\end{array}
\right]\, .
		\label{eqn:Lwitheval}
\end{equation}
To compute with $\ipmatrix(\tv)$  without resorting to cofactor expansions, we
express $\ipmatrix(\tv)$ as a rank-one perturbation of a matrix of numbers.
Define
\begin{equation}
\mnoi{E}
=	
	\left[ 
\begin{array}{cccc}
\overline{d}(1)	& \overline{d}(2)	& \cdots	& \overline{d}(\el)\\
\langle \tilde{\phi}_2,\phi_1\rangle
	&\langle \tilde{\phi}_2,\phi_2\rangle
	& \cdots
	&\langle \tilde{\phi}_2,\phi_{\el}\rangle\\
\vdots	&\vdots 	&\ddots			&\vdots\\
\langle \tilde{\phi}_{\el},\phi_1\rangle
	&\langle \tilde{\phi}_{\el},\phi_2\rangle
	& \cdots
	&\langle \tilde{\phi}_{\el},\phi_{\el}\rangle
\end{array}
\right]\, ,
		\label{eqn:Edef}
\end{equation}
where the vector $\vnoi{d}^*$ is chosen to be a unit vector orthogonal
to the remaining rows of $\mnoi{E}$. This choice assures that the rank
deficiency of $\mnoi{E}$ will be smaller than or equal to the rank
deficiency of the matrix with any other first row. It also gives us
some convenient properties, namely $\mnoi{E}\vnoi{d}=\vnoi{e}_1$,
$\vnoi{d}^*\mnoi{E}^{\ddag}=\vnoi{e}_1^*$,
$\mnoi{E}^{\ddag}\vnoi{e}_1=\vnoi{d}$, and
$\vnoi{e}_1^*\mnoi{E}=\vnoi{d}^*$, where $\mnoi{E}^\ddag$ is the
modified pseudo-inverse of Definition~\ref{def:Addag}.  It costs
$\bigoh(\el^2\res)$ to construct $\mnoi{E}$ and $\bigoh(\el^3)$ to
compute $\mnoi{E}^{\ddag}$ and $|\mnoi{E}|$.

We then have
\begin{equation}
\ipmatrix(\tv)=\mnoi{E}+\vnoi{e}_1(\vnoi{\Phi}(\tv)-\vnoi{d})^*
		\label{eqn:updateL}
\end{equation}
and, with the help of Proposition~\ref{pro:pertrankdet}, compute
\begin{equation}
|\ipmatrix(\tv)|
=
\left|\mnoi{E}\|\mnoi{I}+\vnoi{d}(\vnoi{\Phi}(\tv)-\vnoi{d})^*\right|
=|\mnoi{E}| \left(1+(\vnoi{\Phi}(\tv)-\vnoi{d})^*\vnoi{d}\right)
=|\mnoi{E}|\, \vnoi{\Phi}(\tv)^*\vnoi{d}
\,,
		\label{eqn:updatedetL}
\end{equation}
which yields 
\begin{prop}
\begin{equation}
\left\langle \eval{\tv-\tv_1}\prod_{i=2}^\el \tilde{\phi}_i(\tv_i),
\prod_{i=1}^\el \phi_i(\tv_i) \right\rangle_{\op{A}}
= |\mnoi{E}|\, \vnoi{\Phi}(\tv)^*\vnoi{d}
\,,
		\label{eqn:Aipeval}
\end{equation}
where $\mnoi{E}$ and $\vnoi{d}$ are defined as above.
\end{prop}

\begin{rem}
If $i>1$ then
\begin{equation}
\langle|\mnoi{E}|\, \vnoi{\Phi}^*\vnoi{d}, \tilde{\phi}_i \rangle
=|\mnoi{E}| \langle \vnoi{\Phi}, \tilde{\phi}_i \rangle^*\vnoi{d}
=|\mnoi{E}| E(i,\cdot)^*\vnoi{d}
=0
\,,
\end{equation}
since $\vnoi{d}$ is orthogonal to $E(i,\cdot)$, which is row number
$i$ of $\mnoi{E}$.  
Thus 
the function (\ref{eqn:Aipeval}) is orthogonal to $\tilde{\phi}_i$ for $i>1$.
The same property will hold when the operators $\op{T}$, $\op{V}$, and
$\op{W}$ are present in the antisymmetric inner product, as described
in the following sections.
\end{rem}

\subsection{Antisymmetric Inner Product with $\eval{\tv-\tv_1}$ and
($\op{T}$ and/or $\op{V}$) Present}
\label{sec:AipVeval}

To compute antisymmetric inner products involving operators, we will
modify formulas from Section~\ref{sec:Aip}. The first (trivial)
modification is to denote the variable of integration in those
formulas by $\tv'$, so as not to confuse it with the variable $\tv$ in
$\eval{\tv-\tv_1}$. Next we replace $|\ipmatrix|$ with
$|\ipmatrix(\tv)|$ given by (\ref{eqn:updatedetL}). 
Using (\ref{eqn:updateL}), we can express
\begin{multline}
\ipmatrix(\tv)^{-1}
=\left(\mnoi{E}+\vnoi{e}_1(\vnoi{\Phi}(\tv)-\vnoi{d})^*\right)^{-1}
=\left(\mnoi{E}\left(\mnoi{I}+\vnoi{d}(\vnoi{\Phi}(\tv)-\vnoi{d})^*\right)\right))^{-1}
\\
=\left(\mnoi{I}+\vnoi{d}(\vnoi{\Phi}(\tv)-\vnoi{d})^*\right)^{-1}\mnoi{E}^{-1}
\,.
\end{multline}
Using the Sherman-Morrisson Formula (see e.g.\
\cite{GOL-LOA:1996} and (\ref{eqn:likeshermanmorrison}) in Appendix~\ref{sec:pseudoupdate})
we then have
\begin{equation}
\ipmatrix(\tv)^{-1}
=
\left(\mnoi{I}
-
\frac{\vnoi{d}(\vnoi{\Phi}(\tv)-\vnoi{d})^*}
{1+(\vnoi{\Phi}(\tv)-\vnoi{d})^*\vnoi{d}}\right)
\mnoi{E}^{-1}
=
\left(\mnoi{I}
+
\vnoi{d}
\frac{(\vnoi{d}-\vnoi{\Phi}(\tv))^*}
	{\vnoi{\Phi}(\tv)^*\vnoi{d}}\right)
\mnoi{E}^{-1}
	\label{eqn:updateLinv}
\,.
\end{equation}
The vector of functions $\vnoi{\Theta}$, which was defined by
$\ipmatrix^{-1}\tilde{\vnoi{\Phi}}$, now
depends on the variable $\tv$ in $\eval{\tv-\tv_1}$ as well as its own
internal variable $\tv'$.
Replacing $\ipmatrix^{-1}$ with (\ref{eqn:updateLinv}) and
$\tilde{\vnoi{\Phi}}$ with 
$\tilde{\vnoi{\Phi}}(\tv')+\vnoi{e}_1(\eval{\tv-\tv'}-\tilde{\phi}_1(\tv'))$,
we obtain
\begin{equation}
\vnoi{\Theta}(\tv,\tv')=
\left(\mnoi{I}
+
\vnoi{d}
\frac{(\vnoi{d}-\vnoi{\Phi}(\tv))^*}
	{\vnoi{\Phi}(\tv)^*\vnoi{d}}\right)
\mnoi{E}^{-1}
\left(\tilde{\vnoi{\Phi}}(\tv')+\vnoi{e}_1(\eval{\tv-\tv'}-\tilde{\phi}_1(\tv'))\right)
\,.
		\label{eqn:updateThetaraw}
\end{equation}
To compute it, we first compute the base case
$\tilde{\vnoi{\Theta}}(\tv')=\mnoi{E}^{-1}\tilde{\vnoi{\Phi}}(\tv')$.  
Multiplying out (\ref{eqn:updateThetaraw})
and noting 
$\vnoi{d}^*\tilde{\vnoi{\Theta}}=\vnoi{d}^*\mnoi{E}^{\ddag}\tilde{\vnoi{\Phi}}=\tilde{\phi}_1$,
we obtain
\begin{multline}
\vnoi{\Theta}(\tv,\tv')
=
\tilde{\vnoi{\Theta}}(\tv')
+
\vnoi{d}
\frac{\vnoi{d}^*\tilde{\vnoi{\Theta}}(\tv')
	-\vnoi{\Phi}(\tv)^*\tilde{\vnoi{\Theta}}(\tv')
	+\eval{\tv-\tv'}-\tilde{\phi}_1(\tv')}
	{\vnoi{\Phi}(\tv)^*\vnoi{d}}
\\
=
\tilde{\vnoi{\Theta}}(\tv')
-
\vnoi{d}
\frac{\vnoi{\Phi}(\tv)^*\tilde{\vnoi{\Theta}}(\tv')-\eval{\tv-\tv'}}
	{\vnoi{\Phi}(\tv)^*\vnoi{d}}
\, .
		\label{eqn:updateTheta}
\end{multline}
We are now ready to state our main formulas.
\begin{prop}
When $\mnoi{E}$ is nonsingular,
\begin{equation}
\left\langle 
  \eval{\tv-\tv_1}
\prod_{i=2}^\el  \tilde{\phi}_i(\tv_i)
,
(\op{T}+\op{V})
\prod_{i=1}^\el \phi_i(\tv_i)
\right\rangle_\op{A}
			\label{eqn:bVpart}
\end{equation}
is equal to
\begin{multline}
\frac{|\mnoi{E}|}{{\el!}}
\left[
\vnoi{\Phi}(\tv)^*
\left(
\vnoi{d}
\int \tvtimes{\vnoi{\Phi}}^*\tilde{\vnoi{\Theta}}
d\tv'
-
\int \tvtimes{\vnoi{\Phi}^*
\vnoi{d}}
\tilde{\vnoi{\Theta}}
d\tv'
\right)
\right.\\
+
\tvtimes{\vnoi{\Phi}^*\vnoi{d}}(\tv)
\big]
\,,
		\label{eqn:bAipVresult}
\end{multline}
which can be computed with total cost $\bigoh(\el^3+\el^2\res)$.
		\label{pro:bAipVresult}
\end{prop}
\begin{pf}
To compute (\ref{eqn:bVpart}), we start with $\frac{|\ipmatrix|}{{\el!}}
\int \tvtimes{\vnoi{\Phi}}^*\vnoi{\Theta} d\tv'$ from
(\ref{eqn:AipTVresult}) and 
substitute in (\ref{eqn:updatedetL}) and (\ref{eqn:updateTheta}) 
to obtain
\begin{equation}
\frac{|\mnoi{E}|\, \vnoi{\Phi}(\tv)^*\vnoi{d}}{{\el!}}
\int \tvtimes{\vnoi{\Phi}}(\tv')^*
\left(\tilde{\vnoi{\Theta}}(\tv')
-
\vnoi{d}
\frac{\vnoi{\Phi}(\tv)^*\tilde{\vnoi{\Theta}}(\tv')-\eval{\tv-\tv'}}
	{\vnoi{\Phi}(\tv)^*\vnoi{d}}
\right)
d \tv'
\,.
\end{equation}
Distributing out and
rearranging, we have
\begin{multline}
\frac{|\mnoi{E}|}{{\el!}}
\int
\vnoi{\Phi}(\tv)^*\vnoi{d}
\tvtimes{\vnoi{\Phi}}^*(\tv')
\tilde{\vnoi{\Theta}}(\tv')
-
\tvtimes{
\vnoi{\Phi}}(\tv')^*\vnoi{d}
\vnoi{\Phi}(\tv)^*\tilde{\vnoi{\Theta}}(\tv')
\\
+
\tvtimes{
\vnoi{\Phi}}(\tv')^*\vnoi{d}
\eval{\tv-\tv'}
d\tv'
\,,
\end{multline}
which yields
(\ref{eqn:bAipVresult}).
Although in (\ref{eqn:updateLinv}) and (\ref{eqn:updateTheta}) we divide
by $\vnoi{\Phi}^*\vnoi{d}$, which could be zero, this denominator cancels in the final
expression, so we can argue by continuity that the final expression is
still valid. One can also prove this directly by determining the
nullspace of $\ipmatrix$ and then using (\ref{eqn:AipVQ1final}).
\qed\end{pf}

\begin{rem}
It is the term with pointwise multiplication, $\tvtimes{\vnoi{\Phi}^*\vnoi{d}}$ in
(\ref{eqn:bAipVresult}), that allows adaptive refinement around the
nuclei in the numerical algorithm.
\end{rem}

To obtain the formulas when $\mnoi{E}$ is singular, we follow the same
logic as in Section~\ref{sec:AipWsingular}.  Denote the singular
vectors in the nullspace of $\mnoi{E}$ by
$\{(\tilde{\vnoi{u}}_i,\tilde{\vnoi{v}}_i)\}$.

\begin{prop}
When $\mnoi{E}$ has rank deficiency greater than one, (\ref{eqn:bVpart}) is zero.
When $\mnoi{E}$ has rank deficiency one, (\ref{eqn:bVpart}) is equal to
\begin{equation}
\frac{1}{|\mnoi{E}^{\ddag}|{\el!}}
 \vnoi{\Phi}(\tv)^*
\left(
\vnoi{d}
\int \tvtimes{\vnoi{\Phi}^*\tilde{\vnoi{v}}_1}\tilde{\vnoi{u}}_1^*\tilde{\vnoi{\Phi}}d\tv'
-
\tilde{\vnoi{v}}_1
\int\tvtimes{\vnoi{\Phi}^*\vnoi{d}}
\tilde{\vnoi{u}}_1^*\tilde{\vnoi{\Phi}}d\tv'
\right)
\,,
	\label{eqn:AipevalVQ1}
\end{equation}
which can be computed with total cost
$\bigoh(\el^3+\el^2\res)$.
\end{prop}

\subsection{Antisymmetric Inner Product with $\eval{\tv-\tv_1}$ and $\op{W}$ Present}
\label{sec:AipWeval}

Conceptually the derivation if $\op{W}$ is present in the inner
product is the same and we obtain the
following propositions.
\begin{prop}
When $\mnoi{E}$ is nonsingular,
\begin{equation}
\left\langle 
\eval{\tv-\tv_1}
\prod_{i=2}^\el \tilde{\phi}_i(\tv_i)
,
\op{W}
\prod_{i=1}^\el \phi_i(\tv_i)
\right\rangle_{\op{A}}
			\label{eqn:bWpart}
\end{equation}
is equal to
\begin{multline}
\frac{1}{2}\frac{|\mnoi{E}|}{{\el!}}
\left[
2
\left(
\vnoi{\Phi}(\tv)^*\vnoi{d}\wapply{\vnoi{\Phi}^*\tilde{\vnoi{\Theta}}}(\tv)
	-\vnoi{\Phi}(\tv)^*\wapply{\tilde{\vnoi{\Theta}}\vnoi{\Phi}^*\vnoi{d}}(\tv)
\right)
\right.
\\
+
\vnoi{\Phi}(\tv)^*
\left(
\vnoi{d}
\int 
	\vnoi{\Phi}^*\tilde{\vnoi{\Theta}}
	\wapply{\vnoi{\Phi}^*\tilde{\vnoi{\Theta}}}
	-\vnoi{\Phi}^*
	\wapply{\tilde{\vnoi{\Theta}}\vnoi{\Phi}^*}
	\tilde{\vnoi{\Theta}}
d\tv'
\right.
\\
\left.
\left.
-2
\int
\tilde{\vnoi{\Theta}}
\wapply{\vnoi{\Phi}^*\tilde{\vnoi{\Theta}}}\vnoi{\Phi}^*\vnoi{d}
	-\tilde{\vnoi{\Theta}}
	\vnoi{\Phi}^*\wapply{\tilde{\vnoi{\Theta}}\vnoi{\Phi}^*\vnoi{d}}
d\tv'
\right)
\right]
\,,
	\label{eqn:bAipWresult}
\end{multline}
which can be computed with total cost $\bigoh(\el^3+\el^2\res\log\res)$.
\end{prop}

\begin{prop}
When $\mnoi{E}$ has rank deficiency one, (\ref{eqn:bWpart}) is equal to
\begin{align}
\frac{1}{|\mnoi{E}^{\ddag}|{\el!}}
\Big[
&\left(
\vnoi{\Phi}(\tv)^*\vnoi{d}\wapply{\vnoi{\Phi}^*\tilde{\vnoi{v}}_1\tilde{\vnoi{u}}_1^*\tilde{\vnoi{\Phi}}}(\tv)
-
\vnoi{\Phi}(\tv)^*\tilde{\vnoi{v}}_1\wapply{\tilde{\vnoi{u}}_1^*\tilde{\vnoi{\Phi}}\vnoi{\Phi}^*\vnoi{d}}(\tv)
\right)
\nonumber\\
+
\vnoi{\Phi}(\tv)^*
&\left(
\vnoi{d}\int\vnoi{\Phi}^*\tilde{\vnoi{v}}_1
	\left(\tilde{\vnoi{u}}_1^*\tilde{\vnoi{\Phi}}
	\wapply{\vnoi{\Phi}^*\tilde{\vnoi{\Theta}}}
	-
	\wapply{\tilde{\vnoi{u}}_1^*\tilde{\vnoi{\Phi}}\vnoi{\Phi}^*}
	\tilde{\vnoi{\Theta}}
	\right)d\tv'
\right.
\nonumber\\
&+
\int
\tilde{\vnoi{\Theta}}\left(
\vnoi{\Phi}^*\tilde{\vnoi{v}}_1\wapply{\tilde{\vnoi{u}}_1^*\tilde{\vnoi{\Phi}}\vnoi{\Phi}^*\vnoi{d}}
-
\wapply{\vnoi{\Phi}^*\tilde{\vnoi{v}}_1\tilde{\vnoi{u}}_1^*\tilde{\vnoi{\Phi}}}\vnoi{\Phi}^*\vnoi{d}
\right)
d\tv'
\nonumber\\
&-\left.\left.
\tilde{\vnoi{v}}_1\int\vnoi{\Phi}^*\vnoi{d}
	\left(\tilde{\vnoi{u}}_1^*\tilde{\vnoi{\Phi}}
	\wapply{\vnoi{\Phi}^*\tilde{\vnoi{\Theta}}}
	-\wapply{\tilde{\vnoi{u}}_1^*\tilde{\vnoi{\Phi}}\vnoi{\Phi}^*}\tilde{\vnoi{\Theta}}
	\right)d\tv'
\right)
\right]
\, ,
	\label{eqn:AipWQ11accelerate}
\end{align}
which can be computed with total cost $\bigoh(\el^3+\el^2\res+\el\res\log\res)$.
\end{prop}

\begin{prop}
When $\mnoi{E}$ has rank deficiency two, (\ref{eqn:bWpart}) is equal to
\begin{align}
\frac{1}{|\mnoi{E}^{\ddag}|{\el!}}
\vnoi{\Phi}(\tv)^*
\Big[&
\vnoi{d}
\int 
\vnoi{\Phi}^*\tilde{\vnoi{v}}_1
\tilde{\vnoi{u}}_1^*\tilde{\vnoi{\Phi}}
\wapply{\vnoi{\Phi}^*\tilde{\vnoi{v}}_2\tilde{\vnoi{u}}_2^*\tilde{\vnoi{\Phi}}}
-
\vnoi{\Phi}^*\tilde{\vnoi{v}}_2
\wapply{\tilde{\vnoi{u}}_2^*\tilde{\vnoi{\Phi}}\vnoi{\Phi}^*\tilde{\vnoi{v}}_1}
\tilde{\vnoi{u}}_1^*\tilde{\vnoi{\Phi}}
d\tv\nonumber\\
&-\tilde{\vnoi{v}}_1\int 
\vnoi{\Phi}^*\tilde{\vnoi{v}}_2\tilde{\vnoi{u}}_2^*\tilde{\vnoi{\Phi}}
\wapply{\vnoi{\Phi}^*\vnoi{d}\tilde{\vnoi{u}}_1^*\tilde{\vnoi{\Phi}}}
-
\vnoi{\Phi}^*\tilde{\vnoi{v}}_2
\wapply{\tilde{\vnoi{u}}_2^*\tilde{\vnoi{\Phi}}\vnoi{\Phi}^*\vnoi{d}}
\tilde{\vnoi{u}}_1^*\tilde{\vnoi{\Phi}}
d\tv\nonumber\\ 
&-\tilde{\vnoi{v}}_2\int 
\vnoi{\Phi}^*\tilde{\vnoi{v}}_1
\tilde{\vnoi{u}}_1^*\tilde{\vnoi{\Phi}}
\wapply{\vnoi{\Phi}^*\vnoi{d}\tilde{\vnoi{u}}_2^*\tilde{\vnoi{\Phi}}}
-
\vnoi{\Phi}^*\tilde{\vnoi{v}}_1
\wapply{\tilde{\vnoi{u}}_1^*\tilde{\vnoi{\Phi}}\vnoi{\Phi}^*\vnoi{d}}
\tilde{\vnoi{u}}_2^*\tilde{\vnoi{\Phi}}
d\tv
\,\Big],
	\label{eqn:AipWQ22accelerate}
\end{align}
which can be computed with total cost $\bigoh(\el^3+\el\res+\res\log\res)$.
	\label{pro:bAipWQ2}
\end{prop}


\section{Details of the Green's Function Iteration}
\label{sec:greendetail}

In this section we fill in the missing pieces in the Green's function iteration
algorithm outlined in Section~\ref{sec:maindescription}. First we
give a representation for
the Green's function itself. Then we use the methods in the previous
sections to construct the vector $\vnoi{b}$ in (\ref{eqn:bdefplain})
and the matrix $\mnoi{A}$ in (\ref{eqn:normalAdef}) to form the normal
equations (\ref{eqn:normal}). Next we give the algorithm from
Section~\ref{sec:maindescription} in outline form as pseudocode.
Finally we gather the computational cost
of the whole method, and present some linear algebra techniques to
reduce it.

\subsection{Representing the Green's Function}
\label{sec:greenrepresent}

In this section we construct a separated representation for the
Green's function $\op{G}_\mu$ in (\ref{eqn:Gmudef}), following the ideas in \cite{BEY-MOH:2002,BEY-MOH:2005} 
(see also \cite{HAC-KHO:2006,HAC-KHO:2006a}). 
We will use this representation in Section~\ref{sec:constructb} when
constructing the right-hand-side of the normal equations.

We begin by constructing an approximation of $1/t$ with exponentials
such that
\begin{equation}
\left|
\frac{1}{t}-\sum_{p=1}^{L} w_p \exp({-\tau_p t})
\right| < \epsilon
\,,
\label{eqn:1overtapprox}
\end{equation}
on the interval $t\in [1,\infty)$, with $w_p$ and $\tau_p$
positive. Expansions of $1/t$ into exponentials have been used in
several applications and constructed by diverse techniques; see
\cite{BRAESS:1995,HRY-ROK:1998,YAR-ROK:1999,BEY-MON:2005,BRA-HAC:2005,HACKBU:2005T}
and the references therein. The interval $[1,\infty)$ is addressed
specifically in \cite{BRA-HAC:2005}, where it is shown that the error
rate $\epsilon=\bigoh(\exp({-c\sqrt{L}}))$ can be achieved, which
means we can achieve $L=\bigoh((\ln \epsilon)^2)$.

Substituting $t=s/(-\mu)$ for $\mu<0$ into (\ref{eqn:1overtapprox})
and dividing by $-\mu$, one has
\begin{equation}
\left|
\frac{1}{s}-\sum_{p=1}^{L} \frac{w_p}{-\mu} \exp({-\frac{\tau_p}{-\mu} s})
\right| < \frac{\epsilon}{-\mu}
\,,
\label{eqn:1overtapproxMU}
\end{equation}
valid on the interval $s\in [-\mu,\infty)$. In Fourier coordinates, we
can express
\begin{equation}
\op{G}_\mu =
\frac{1}{2 \pi^2 \sum \xi_i^2 -\mu}\,, 
\end{equation}
from which we see that $\|\op{G}_\mu\|=1/(-\mu)$. Since the
denominator is at least $-\mu>0$, we can substitute into
(\ref{eqn:1overtapproxMU}) and obtain
\begin{equation}
\left|
\op{G}_\mu
-\sum_{p=1}^{L} \frac{w_p}{-\mu} e^{-\tau_p} 
\bigotimes_{i=1}^\el \exp({-\frac{2\pi^2 \tau_p}{-\mu} \xi_i^2})
\right| < \frac{\epsilon}{-\mu}=\epsilon\|\op{G}_\mu\|
\,.
\end{equation}
Thus we obtain an approximation of $\op{G}_\mu$ with relative error
$\epsilon$ in norm using $L$ terms, with $L$ independent of $\el$ and
$\mu$.
To construct $\op{G}_\mu$ as an integral operator in spatial
coordinates, we apply the inverse Fourier transform to obtain
\begin{equation}
\op{G}_\mu
\approx 
\sum_{p=1}^L  
\bigotimes_{i=1}^{\el} \gaussop^p_{\pv_i} \, ,
		\label{eqn:gmuapprox}
\end{equation}
where the convolution operator $\gaussop^p_{\pv_i}$, which depends implicitly on $\mu$, 
is defined by
\begin{multline}
\gaussop^p_{\pv_i} f(\tv_1,\ldots,\tv_\el)
= 
\left(\frac{w_p}{-\mu e^{\tau_p}}\right)^{1/\el}
\left(\frac{-\mu}{2 \pi \tau_p}\right)^{3/2}
\times\\
\int 
\exp\left(-\frac{-\mu}{2\tau_p} {\|\pv_i-\pv'\|^2}\right)
 	f(\tv_1,\ldots,\tv_{i-1},(\pv',\sigma_i),\tv_{i+1},\ldots,\tv_\el) d\pv' \, .
							\label{eqn:Glidef}
\end{multline}


This construction has theoretical value, since it has proved the following theorem.
\begin{thm} 
For any $\epsilon>0$, $\mu<0$, and $\el$, the $\el$-particle Green's
function $\op{G}_\mu$ has a separated representation with relative
error in operator norm bounded by $\epsilon$ using $L=\bigoh((\ln
\epsilon)^2)$ terms, with $L$ independent of $\mu$ and $\el$.
\label{thm:greenrank}
\end{thm}

\subsection{Constructing the Right-Hand-Side Vector $\vnoi{b}$ in (\ref{eqn:bdefplain})}
\label{sec:constructb}

In order to do a step in the iteration, we need to construct the
right-hand-side $\vnoi{b}$ in the normal equations (\ref{eqn:normal})
in Section~\ref{sec:greenproject}.
Since $\op{A}$ is an orthogonal projection,
$\op{A}$ and $\op{G}_\mu$ commute, and $\op{G}_\mu$ is
self-adjoint, the entry (\ref{eqn:bdefplain}) is equal to
\begin{equation}
b(l)(\tv)
=-\tilde{s}_l\sum_{m}^{\srank} \s_{m}  
\left\langle 
{\op{A}}\,\op{G}_{\mu}
\eval{\tv-\tv_1}
\prod_{i=2}^\el \tilde{\phi}_i^{l}(\tv_i)
,
[\op{V}+\op{W}]
\prod_{i=1}^\el \phi_i^{m}(\tv_i)
\right\rangle
\,.
\end{equation}
Substituting (\ref{eqn:gmuapprox}) in for $\op{G}_\mu$ and rearranging, we have
\begin{equation}
b(l)(\tv)
=-\tilde{s}_l\sum_{m}^{\srank} \s_{m}  
\sum_{p=1}^L  
\left\langle 
{\op{A}}
\gaussop^p_{\pv_1} \eval{\tv-\tv_1}
\prod_{i=2}^\el \gaussop^p_{\pv_i} \tilde{\phi}_i^{l}(\tv_i)
,
[\op{V}+\op{W}]
\prod_{i=1}^\el \phi_i^{m}(\tv_i)
\right\rangle
\,.
			\label{eqn:bGexpanded}
\end{equation}
The computation is of the same form for each value of the
indices $l$, $m$, and $p$, so we can consider a single term and suppress the indices.
%

To evaluate a single term $\langle 
{\op{A}}
\gaussop_{\pv_1} \eval{\tv-\tv_1}
\prod_{i=2} \gaussop_{\pv_i} \tilde{\phi}_i(\tv_i)
,
[\op{V}+\op{W}]
\prod_{i=1} \phi_i(\tv_i)
\rangle$
we use the formulas in
Propositions~\ref{pro:bAipVresult}--\ref{pro:bAipWQ2}
in Sections~\ref{sec:AipVeval} and~\ref{sec:AipWeval}, with two
modifications.  The first modification is that $\tilde{\vnoi{\Phi}}$
is replaced with $\gaussop\tilde{\vnoi{\Phi}}$ throughout. This
replacement causes no structural change to the formulas; it just
changes the inputs.
The second
modification is caused by the replacement of $\eval{\tv-\tv_1}$ by
$\gaussop_{\pv_1}\eval{\tv-\tv_1}$. 
The first row of $\ipmatrix(\tv)$ in
(\ref{eqn:Lwitheval}) becomes $\gaussop\vnoi{\Phi}(\tv)^*$, which makes
$|\ipmatrix(\tv)|= |\mnoi{E}|\,\gaussop\vnoi{\Phi}(\tv)^*\vnoi{d}$.  
Similarly, (\ref{eqn:updateTheta}) becomes 
\begin{equation}
\vnoi{\Theta}(\tv,\tv')
=
\tilde{\vnoi{\Theta}}(\tv')
-
\vnoi{d}
\frac{\gaussop\vnoi{\Phi}(\tv)^*\tilde{\vnoi{\Theta}}(\tv')-\gaussop\eval{\tv-\tv'}}
	{\gaussop\vnoi{\Phi}(\tv)^*\vnoi{d}}
\, .
\end{equation}
Tracking $\gaussop$ through the formulas, we find that all
we need to do is to modify the formulas in Sections~\ref{sec:AipVeval} and~\ref{sec:AipWeval}
by applying $\gaussop$ to the final result.

\subsection{Constructing the Matrix $\mnoi{A}$ in (\ref{eqn:normalAdef})}
\label{sec:constructA}

In this section we construct the kernels in (\ref{eqn:normalAdef}) for the normal equations (\ref{eqn:normal}),
using the same ideas as in Section~\ref{sec:Aipeval}.
We fix $l$ and $l'$ and define
\begin{eqnarray}
K(\tv,\tv')	& =	&\frac{A(l,l')(\tv,\tv')}{\tilde{s}_l\tilde{s}_{l'}}\\
\vnoi{w}(\tv')	&=	&
	\left[\begin{array}{ccc}
	\tilde{\phi}^{l}_2(\tv')	&\hdots & \tilde{\phi}^{l}_\el(\tv')
	      \end{array}\right]^*\\
\vnoi{y}(\tv)	&=	&
	\left[\begin{array}{ccc}
	\tilde{\phi}^{l'}_2(\tv)	&\hdots & \tilde{\phi}^{l'}_\el(\tv)
	      \end{array}\right]^*\\
\mnoi{D}	&=	&
	\left[ 
\begin{array}{ccc}
\langle \tilde{\phi}^{l}_2,\tilde{\phi}^{l'}_2\rangle
	   &\cdots	&\langle \tilde{\phi}^{l}_2,\tilde{\phi}^{l'}_\el\rangle\\
\vdots 	&\ddots			&\vdots\\
\langle \tilde{\phi}^{l}_\el,\tilde{\phi}^{l'}_2\rangle
	   &\cdots  	&\langle \tilde{\phi}^{l}_\el,\tilde{\phi}^{l'}_\el\rangle
\end{array}
\right]
		\label{eqn:DdefforA}
\, .
\end{eqnarray}
Using L\"owdin's rules (\ref{eqn:lowdin}) we have 
\begin{equation}
K(\tv,\tv')=\frac{|\ipmatrix|}{{\el!}}=
\frac{1}{{\el!}}
\left|\begin{array}{cc}  
\delta(\tv-\tv')	& \vnoi{y}^*(\tv) \\
\vnoi{w}(\tv')	& \mnoi{D}
      \end{array}
\right|
\,.
	\label{eqn:Akernelbasic}
\end{equation}
Expressing $\ipmatrix$ as a low-rank perturbation of $\left[\begin{array}{cc}  
1	& 0 \\
0	& \mnoi{D}
      \end{array}
\right]$, 
we have
\begin{multline}
K(\tv,\tv')=
\frac{1}{{\el!}}\left|
\left[\begin{array}{cc}  
1	& 0 \\
0	& \mnoi{D}
      \end{array}
\right]
+
\left[ \begin{array}{c} 1 \\ 0 \end{array} \right]
\left[ \begin{array}{cc} 0 & \vnoi{y}^*(\tv) \end{array}\right]
+
\left[\begin{array}{c} \delta(\tv-\tv')-1\\ 
\vnoi{w}(\tv') \end{array} \right] 
\left[ \begin{array}{cc} 1 & 0 \end{array}\right]
\right|\\
=
\frac{1}{{\el!}}\left|\begin{array}{cc}  
1	& 0 \\
0	& \mnoi{D}
      \end{array}
\right|
\left|
\mnoi{I} +
\left[ \begin{array}{c} 1 \\ 0 \end{array} \right]
\left[ \begin{array}{cc} 0 & \vnoi{y}^*(\tv) \end{array}\right]
+
\left[\begin{array}{c} \delta(\tv-\tv')-1\\ 
\mnoi{D}^{-1}\vnoi{w}(\tv') \end{array} \right] 
\left[ \begin{array}{cc} 1 & 0 \end{array}\right]
\right|\\
=
\frac{\left|\mnoi{D}\right|}{{\el!}}
\left|
\begin{array}{cc}
1	&\vnoi{y}^*(\tv) \mnoi{D}^{-1}\vnoi{w}(\tv') \\
1	&\delta(\tv-\tv')
\end{array}
\right|
=
\frac{\left|\mnoi{D}\right|}{{\el!}}
\left(\delta(\tv-\tv')-\vnoi{y}^*(\tv) \mnoi{D}^{-1}\vnoi{w}(\tv') \right)
\, .
	\label{eqn:Aconstruct}
\end{multline}

If $\mnoi{D}$ is singular then we apply the same logic as in
Section~\ref{sec:AipWsingular}. If $\mnoi{D}$ has rank-deficiency
greater than one then $K(\tv,\tv')=0$.  If it has rank-deficiency one
then we have $K(\tv,\tv')=$
\begin{multline}
\frac{1}{|\mnoi{D}^{\ddag}|{\el!}}
\left|
\mnoi{I} +
\left[ \begin{array}{c} 0 \\ -\vnoi{v} \end{array} \right]
\left[ \begin{array}{cc} 0 & \vnoi{v}^* \end{array}\right]
+
\left[ \begin{array}{c} 1 \\ 0 \end{array} \right]
\left[ \begin{array}{cc} 0 & \vnoi{y}^*(\tv) \end{array}\right]
+
\left[\begin{array}{c} \delta(\tv-\tv')-1\\ 
\mnoi{D}^{\ddag}\vnoi{w}(\tv') \end{array} \right] 
\left[ \begin{array}{cc} 1 & 0 \end{array}\right]
\right|\\
=
\frac{1}{|\mnoi{D}^{\ddag}|{\el!}}
\left|
\begin{array}{ccc}
0			&0	&\vnoi{v}^* \mnoi{D}^{\ddag}\vnoi{w}(\tv') \\
-\vnoi{y}^*(\tv)\vnoi{v}	&1	&\vnoi{y}^*(\tv) \mnoi{D}^{\ddag}\vnoi{w}(\tv') \\
0			&1	&\delta(\tv-\tv')
\end{array}
\right|
=
\frac{-(\vnoi{y}^*(\tv)\vnoi{v})(\vnoi{v}^*\mnoi{D}^{\ddag}\vnoi{w}(\tv'))}
{|\mnoi{D}^{\ddag}|{\el!}}
\\
=
\frac{-(\vnoi{y}^*(\tv)\vnoi{v})(\vnoi{u}^*\vnoi{w}(\tv'))}
{|\mnoi{D}^{\ddag}|{\el!}}
\, ,
\label{eqn:accelfactoredQ1}
\end{multline}
where $\mnoi{D}^{\ddag}$ is the
modified pseudo-inverse of Definition~\ref{def:Addag}.

In the nonsingular case, we can construct $\mnoi{D}$ at cost
$\bigoh(\el^2\res)$ and compute $\mnoi{D}^{-1}$ at cost
$\bigoh(\el^3)$.  Applying this kernel costs $\bigoh(\el\res)$ to
integrate against a function in $\tv'$, $\bigoh(\el^2)$ to apply
$\mnoi{D}^{-1}$, and then $\bigoh(\el\res)$ to apply $\vnoi{y}^*$ to
the result.
In the singular case,
we can compute $\mnoi{D}^{\ddag}$ at cost $\bigoh(\el^3)$ and
construct $\vnoi{y}^*\vnoi{v}$ and $\vnoi{u}^*\vnoi{w}$ at cost
$\bigoh(\el \res)$. Since the variables separate, applying this
kernel costs $\bigoh(\res)$.

\begin{rem}
In the case $\srank=1$, which corresponds to the Hartree-Fock formulation, $\mnoi{D}=\mnoi{I}$ and $K(\tv,\tv')$ 
is just the projector orthogonal to $\{\tilde{\phi}_i\}_{i=2}^\el$.
\end{rem}
 
\subsection{Pseudocode}

In this section we give the algorithm in outline form as pseudocode. We do
not indicate when objects can be recalled or updated from previous computations.
\begin{tabbing}
Lo\=op  through $I$ Green's function iterations
(\ref{eqn:Greeniter},\ref{eqn:Greeniternorm},\ref{eqn:Greenitermu}).
For each of these:
\+\\
Construct $\op{G}_\mu$ as in Section~\ref{sec:greenrepresent},
obtaining the operators $\gaussop^p$ in (\ref{eqn:Glidef}).\\
Lo\=op through the $\el$ directions (electrons). For each of these:\+\\
Compute $A(l,l')$ via (\ref{eqn:Aconstruct}) for all $(l,l')$.\\
Co\=mpute  $b(l)(\tv)$ in (\ref{eqn:bGexpanded}) by:\+\\
Lo\=op  in the $\srank$ values of $l$ and for each:\+\\
Su\=m  over the $L$ values of $p$ and for each:\+\\
Co\=mpute  
	$\gaussop^p\phi_i^l$ for all $i$.\\
Su\=m  over the $\srank$ values of $m$ and for each:\+\\
Using $\gaussop^p\tilde{\vnoi{\Phi}}$ in place of $\tilde{\vnoi{\Phi}}$,
construct $\mnoi{E}$ in (\ref{eqn:Edef}).\\
Compute $|\mnoi{E}|$ and $\mnoi{E}^{-1}$.\\
Construct
$\tilde{\vnoi{\Theta}}=\mnoi{E}^{-1}\gaussop\tilde{\vnoi{\Phi}}$.\\ 
Construct $\vnoi{\Phi}^*\tilde{\vnoi{\Theta}}$,
$\vnoi{\Phi}^*\vnoi{d}$, and $\tilde{\vnoi{\Theta}}\vnoi{\Phi}^*$.\\
Compute $\wapply{\vnoi{\Phi}^*\tilde{\vnoi{\Theta}}}$ and $\wapply{\tilde{\vnoi{\Theta}}\vnoi{\Phi}^*}$.\\
Compute (\ref{eqn:bAipVresult}) and (\ref{eqn:bAipWresult}) using
these ingredients.\\
Apply $\gaussop^p$ to $((\ref{eqn:bAipVresult})+(\ref{eqn:bAipWresult}))$.
\-\-\-\-\\
Apply conjugate gradient to solve the normal equations
(\ref{eqn:normal}).\-\\
Renormalize as in (\ref{eqn:Greeniternorm}).\\
Update $\mu$ via (\ref{eqn:Greenitermu}).
\end{tabbing}

\begin{rem}
We have presented the algorithm in serial form for clarity. The loop
in $l$, sum in $p$, and sum in $m$ can be trivially
parallelized. Parallelizing the loop through the $\el$ electrons would
represent a change in  the algorithm, which we will develop elsewhere. 
\end{rem}

\subsection{Overall Computational Cost}
\label{sec:greencost}

The computational cost is dominated by the repeated construction and
solution of the normal equations (\ref{eqn:normal}).
For a fixed direction, the construction cost is dominated by
(\ref{eqn:bGexpanded}), which has $\srank^2 L$ inner products. The
most costly portion of the inner products is (\ref{eqn:bAipWresult}),
which requires $\bigoh(\el^3+\el^2\res\log\res)$ operations, giving us the net
construction cost 
\begin{equation}
\bigoh(\srank^2 L \el^2(\el +\res\log\res))
\,.
\end{equation}
The operation count to solve the normal equations (\ref{eqn:normal})
by applying the matrix of integral operators $\mnoi{A}$ $S$ times is 
\begin{equation}
\bigoh(\srank^2 S \el(\el +\res))
\,.
\end{equation}

As we loop through the directions, we may reuse several quantities, so
the total cost of the construction is less than $\el$ times the cost
for one direction. In fact, the construction cost for the entire loop
through $\el$ directions is of the same order as the cost for one
direction. The application cost is simply multiplied by $\el$. In the
sections below we show how to update the construction for direction
$k=2$ using what we already have for direction $k=1$, and then
determine the cost for one loop through the directions.  We defer the
development of the technical linear algebra rules on low-rank updates
to Appendix~\ref{sec:pseudoupdate}, and here only show how to apply
them to our problem. Our final conclusion is the computational cost
\begin{equation}
\bigoh(I \srank^2 \el^2
[
L(\el+\res\log\res)
+S(\el+\res)
]
),
\end{equation}
where $I$ the number of Green's function iterations.

\subsubsection{Reuse in Computing $\mnoi{A}$}
\label{sec:Areuse}

Let $\mnoi{D}_1$ denote $\mnoi{D}$ in (\ref{eqn:DdefforA}) for
directions one, and $\mnoi{D}_2$ the version for direction two.
We let $\hat{\phi}^{l}_1$ denote the updated version of
$\tilde{\phi}^{l}_1$.  
To construct $\mnoi{D}_2$ requires only the first column and row of
$\mnoi{D}_1$ to be updated, specifically
\begin{equation}
\mnoi{D}_2=\mnoi{D}_1
+\vnoi{e}_1 \left[ \begin{array}{ccc}
	0	
	&(\langle\hat{\phi}^{l}_1,\tilde{\phi}^{l'}_3\rangle
		-\langle\tilde{\phi}^{l}_2,\tilde{\phi}^{l'}_3\rangle)
	&\hdots\\  \end{array}\right]
+\left[\begin{array}{c}
	\langle\hat{\phi}^{l}_1,\hat{\phi}^{l'}_1\rangle
		-\langle\tilde{\phi}^{l}_2,\tilde{\phi}^{l'}_2\rangle\\
	\vdots
       \end{array}\right]
	\vnoi{e}_1^*
\,. 
\end{equation}
Computing those inner products involving $\hat{\phi}^{l}_1$
and $\hat{\phi}^{l'}_1$ costs 
$\bigoh(\el\res)$. 
Using Proposition~\ref{pro:pseudoshermanmorrison} twice, we compute
$\mnoi{D}_2^\ddag$, $|\mnoi{D}_2^\ddag|$, and if appropriate $\vnoi{v}$,
all at cost $\bigoh(\el^2)$. 
The formulas (\ref{eqn:Akernelbasic}) and following are modified by
inserting the extra column and row in the second place instead of the
first, but otherwise the procedure is unchanged.
The cost for one loop through the $\el$ directions is thus $\bigoh(\el^3+\el^2\res)$. 


\subsubsection{Reuse in Computing Antisymmetric Inner Products with $\eval{\tv-\tv_1}$ and Operators}
\label{sec:breuse}

We again let $\hat{\phi}^{l}_1$ denote the updated version of
$\tilde{\phi}^{l}_1$ computed during the $k=1$ solve.   
The inner products needed to
construct $\mnoi{E}_2$ require only the one row involving
$\hat{\phi}_1$ to be updated, at cost $\bigoh(\el\res)$.  
The vector $\vnoi{d}_1$ can be constructed by doing the SVD of
$\mnoi{E}_1$ with the first row set to zero and then selecting one of
the right singular vectors $\vnoi{v}_i$ with zero singular value.
Using Proposition~\ref{pro:pseudoshermanmorrison} we obtain
the SVD of $\mnoi{E}_2$ with first row set to zero and second row
containing the new inner products, and thus can find $\vnoi{d}_2$.
Putting the first and second rows back in proper position, we then
have
\begin{multline}
\mnoi{E}_2= \mnoi{E}_1 
+\vnoi{e}_1
\left(\left[ 
\begin{array}{ccc}
\langle \gaussop\hat{\phi}_1,\phi_1\rangle
	& \cdots
	&\langle \gaussop\hat{\phi}_1,\phi_{\el}\rangle
\end{array}\right]
-\vnoi{d}_1^*\right)
\\
+\vnoi{e}_2 \left(\vnoi{d}_2^*-
	\left[ 
\begin{array}{ccc}
\langle \gaussop\tilde{\phi}_2,\phi_1\rangle
	& \cdots
	&\langle \gaussop\tilde{\phi}_2,\phi_{\el}\rangle
\end{array}
\right]\right)\, ,
\end{multline}
and we can compute $|\mnoi{E}_2^\ddag|$ and $\mnoi{E}_2^\ddag$ using
Proposition~\ref{pro:pseudoshermanmorrison} twice, at cost
$\bigoh(\el^2)$. 

Proposition~\ref{pro:pseudoshermanmorrison} produces a rank two
update and we must apply it twice.  For notational ease we will show
how to use a rank one update applied once; the method easily extends.
Assuming $\mnoi{E}^{\ddag}_2=\mnoi{E}^{\ddag}_1+\vnoi{f}\vnoi{g}^*$,
we next update
\begin{multline}
\tilde{\vnoi{\Theta}}_2
=\mnoi{E}^{\ddag}_2\gaussop\tilde{\vnoi{\Phi}}_2
=(\mnoi{E}^{\ddag}_1+\vnoi{f}\vnoi{g}^*)(\gaussop\tilde{\vnoi{\Phi}}_1+\vnoi{e}_1(\hat{\phi}_1-\tilde{\phi}_1))
\\
=\tilde{\vnoi{\Theta}}_1
	+\vnoi{d}_1(\hat{\phi}_1-\tilde{\phi}_1)
	+\vnoi{f}\vnoi{g}^*\gaussop\tilde{\vnoi{\Phi}}_1
	+\vnoi{f}\vnoi{g}^*\vnoi{e}_1(\hat{\phi}_1-\tilde{\phi}_1)
\end{multline}
at cost $\bigoh(\el\res)$.
It is insufficient to just update $\tilde{\vnoi{\Theta}}_2$ in this
way, since it would still cost $\bigoh(\el^2\res\log\res)$ to compute
$\wapply{\tilde{\vnoi{\Theta}}_2\vnoi{\Phi}^*}$ in
(\ref{eqn:bAipWresult}). Instead we update the combined quantity
\begin{multline}
\vnoi{\Phi}^*\wapply{\tilde{\vnoi{\Theta}}_2\vnoi{\Phi}^*}
=
\vnoi{\Phi}^*\wapply{\tilde{\vnoi{\Theta}}_1\vnoi{\Phi}^*}
+\vnoi{\Phi}^*\vnoi{d}_1\wapply{(\hat{\phi}_1-\tilde{\phi}_1)\vnoi{\Phi}^*}
\\
+\vnoi{\Phi}^*\vnoi{f}\wapply{\vnoi{g}^*\gaussop\tilde{\vnoi{\Phi}}_1\vnoi{\Phi}^*}
+\vnoi{\Phi}^*\vnoi{f}\vnoi{g}^*\vnoi{e}_1\wapply{(\hat{\phi}_1-\tilde{\phi}_1)\vnoi{\Phi}}
\end{multline}
at cost $\bigoh(\el\res\log\res)$. With this quantity and
$\tilde{\vnoi{\Theta}}_2$ we can compute (\ref{eqn:bAipWresult}) at
cost $\bigoh(\el\res\log\res)$. The singular cases work similarly.
The cost for one loop through the $\el$ directions is thus $\bigoh(\el^2\res\log\res)$.


\subsection*{Acknowledgments}
We would like to thank Dr.~Robert Harrison (U. of Tennessee and Oak
Ridge National Lab) and Dr.~Lucas Monz\'on (U. of Colorado) for many
useful conversations.  Portions of this research were conducted while
G.B. and M.J.M. were in residence at the Institute for Pure and
Applied Mathematics during the fall of 2004.

This material is based upon work supported by the 
National Science Foundation under Grants
DMS-0219326 (G.B and M.J.M.), 
DMS-0612358 (G.B. and F.P.), 
and DMS-0545895 (M.J.M.), 
the DARPA/ARO under 
Grants W911NF-04-1-0281 
and W911NF-06-1-0254 (G.B and M.J.M.),
and the Department of Energy under Grants 
DE-FG02-03ER25583 and DOE/ORNL Grant 4000038129 (G.B. and F.P.).

\appendix
\renewcommand{\theequation}{\Alph{section}\arabic{equation}}
\setcounter{equation}{0}
\section{Appendix: Algorithms Based on Gradient Descent}
\label{sec:descentdescription}

We prefer the integral iteration in Section~\ref{sec:greendescription}
due to the generally superior numerical properties of integral
formulations. One could, however, try to minimize
(\ref{eqn:variationallambda}) directly with a method based on
gradients. Since the machinery that we have constructed applies to
these methods as well, we sketch how it can be used. 

To minimize (\ref{eqn:variationallambda}) we could use a gradient descent,
starting at some initial guess for $\psi$.
Inserting our current approximation $\psi$ and formally taking the
gradient, we have  
\begin{equation}
2\frac{\aip{\op{H}\psi,\nabla\psi}\aip{\psi,\psi}
-\aip{\op{H}\psi,\psi}\aip{ \psi,\nabla\psi }}
	{\aip{ \psi,\psi}^2}
		\label{eqn:variationalgrad}
\,.
\end{equation}
Defining $\mu$ to be our current value of (\ref{eqn:variationallambda}),
the gradient reduces to
\begin{equation}
\frac{2}{\aip{\psi,\psi}}
\left(\aip{\op{H}\psi,\nabla\psi }
-\mu\aip{\psi,\nabla\psi }
\right)	
		\label{eqn:variationalgradred}
\,.
\end{equation}
The gradient is with respect to the parameters that are used to
minimize (\ref{eqn:variationallambda}). In our case that is the
values of the functions $\phi_j^l$. 
Taking the gradient with respect to the point values of $\phi_j^l$ results in a vector
$\vnoi{g}$ of functions, defined by
\begin{equation}
g_j^l(\tv)
=\frac{2}{\aip{\psi,\psi}}
\s_l \sum_{m=1}^{\srank} \s_{m}  
\aip{
\eval{\tv-\tv_j}
\prod_{i\not=j}^\el \phi_i^{l}(\tv_i)
,
(\op{H}-\mu\op{I})
\prod_{i=1}^\el \phi_i^{m}(\tv_i)
}
\,,
			\label{eqn:gdefvariational}
\end{equation}
where $\eval{\tv-\tv_j}$ is the delta function.
The methods in Section~\ref{sec:Aipeval} can be used to construct $\vnoi{g}$.

Moving $t$ in the direction opposite the gradient replaces $\psi$ with
\begin{equation}
\sum_{l=1}^{\srank} \s_l\prod_{i=1}^\el (\phi_i^l-tg_i^l)
\,.
	\label{eqn:psiupdategradall}
\end{equation}
Some search procedure can then be used to find an appropriate
$t$. Then $\psi$ is updated and the procedure repeated.

Alternatively, we can use an alternating direction approach and take
the gradient with respect to the functions $\phi_i^l$ for one
direction $i$, while fixing the functions in the other directions, and
then loop through the directions.  This loop through the directions is
then repeated $I$ times until we obtain the desired accuracy.  We
describe the $i=1$ case.
Moving $t$ in the direction opposite the gradient replaces $\psi$ with
\begin{equation}
\sum_{l=1}^{\srank}\s_l(\phi_1^l-t g_1^l)\prod_{i=2}^\el \phi_i^{l}
=\psi -t \sum_{l=1}^{\srank}\s_l g_1^l\prod_{i=2}^\el \phi_i^{l}
=\psi -t \tilde{\psi}
\,.
	\label{eqn:psiupdategrad}
\end{equation}
Inserting (\ref{eqn:psiupdategrad}) into (\ref{eqn:variationallambda}) results in
\begin{equation}
\frac{\aip{\op{H}(\psi-t \tilde{\psi}), \psi-t \tilde{\psi} }}
	{\aip{\psi-t \tilde{\psi}, \psi-t \tilde{\psi} }}
=\frac{
\aip{\op{H}\psi, \psi }
-2t\aip{\op{H}\psi,\tilde{\psi} }
+t^2\aip{\op{H}\tilde{\psi},\tilde{\psi} }
}
{\aip{\psi,\psi}-2t\aip{\psi,\tilde{\psi} }
+t^2\aip{\tilde{\psi}, \tilde{\psi} }
}
	\label{eqn:variationalt}
\,.
\end{equation}
Once the inner products have been computed, we can find the minimizer
for (\ref{eqn:variationalt}) by solving a quadratic equation,
and then update $\psi$ via (\ref{eqn:psiupdategrad}).
%
%
%
%
%
%
%
%
The cost to construct $\vnoi{g}$ for one direction is $\srank^2$ times the cost for one
inner product. The dominant cost for the inner product comes from (\ref{eqn:bAipWresult}),
which costs $\bigoh(\el^3+\el^2\res\log\res)$, giving us the net
construction cost 
\begin{equation}
\bigoh(\srank^2 \el^2(\el +\res\log\res))
\,.
\end{equation}
As described in Section~\ref{sec:breuse}, many of the computations can
be reused, so the cost for a single loop through the $\el$ directions
is of the same order.
Thus, for $I$ loops through the directions the overall computational cost is
\begin{equation}
\bigoh(I\srank^2 \el^2(\el +\res\log\res))
\,.
\end{equation}







\setcounter{equation}{0}
\section{Appendix: Low-rank Updates}
\label{sec:pseudoupdate} 

In this section we develop formulas for low-rank updates to 
$\mnoi{A}^\dag$, $\mnoi{A}^\perp$ and $|\mnoi{A}^\ddag|$, based on
\cite{MEYER:1973,BA-BA-TR:2003}. 

\begin{prop}
Given $\mnoi{A}$, $\mnoi{A}^\dag$, $\mnoi{A}^\perp$,
$|\mnoi{A}^\ddag|$, 
$\vnoi{b}$, and $\vnoi{c}$, let
$\mnoi{A}_1=\mnoi{A}+\vnoi{b}\vnoi{c}^*$ and
compute
\begin{equation}
\begin{array}{cccc}
\vnoi{d}=\mnoi{A}^\dag \vnoi{b},
&\vnoi{e}=(\mnoi{A}^\dag)^* \vnoi{c},
&\vnoi{f}=(\mnoi{I}-\mnoi{A}\mnoi{A}^\dag)\vnoi{b},
&\vnoi{g}=(\mnoi{I}-\mnoi{A}^\dag\mnoi{A})\vnoi{c},\\
d=\vnoi{d}^*\vnoi{d},
&e=\vnoi{e}^*\vnoi{e},
&f=\vnoi{f}^*\vnoi{f},
&g=\vnoi{g}^*\vnoi{g},
\\
\lambda=1+\vnoi{c}^*\mnoi{A}^\dag\vnoi{b},
&\mu=|\lambda|^2+dg,
&\nu=|\lambda|^2+ef,\\
\vnoi{p}=\bar{\lambda}\vnoi{d}+d\vnoi{g},
&\vnoi{q}=\lambda\vnoi{e}+e\vnoi{f}.
\end{array}
			\label{eqn:pinvdefs}
\end{equation}
\begin{enumerate}
\item 
If $\lambda=0$, $f=0$, and $g=0$, then $rank(\mnoi{A}_1)=rank(\mnoi{A})-1$ and
\begin{align}
\mnoi{A}_1^{\dag}
&=\mnoi{A}^{\dag}-d^{-1}\vnoi{d}\vnoi{d}^*\mnoi{A}^\dag
+e^{-1}(-\mnoi{A}^\dag\vnoi{e}
+d^{-1}(\vnoi{d}^*\mnoi{A}^\dag\vnoi{e})\vnoi{d})\vnoi{e}^*\\
\mnoi{A}_1^\perp&=\mnoi{A}^\perp +
{(1/\sqrt{de})}{\vnoi{d}\vnoi{e}^*}
	\label{eqn:updateperp1}\\
|\mnoi{A}_1^\ddag|&=-{(1/\sqrt{de})}|\mnoi{A}^\ddag|
	\label{eqn:updatedet1}
\,.
\end{align}
\item 
If $\lambda\not=0$, $f=0$, and $g=0$, then $rank(\mnoi{A}_1)=rank(\mnoi{A})$ and
\begin{align}
\mnoi{A}_1^{\dag}
&=\mnoi{A}^{\dag}-\lambda^{-1}\vnoi{d}\vnoi{e}^*
		\label{eqn:likeshermanmorrison}\\
\mnoi{A}_1^\perp&=\mnoi{A}^\perp
	\label{eqn:updateperp2}\\
|\mnoi{A}_1^\ddag|&=|\mnoi{A}^\ddag|\lambda^{-1}
	\label{eqn:updatedet2}
\,.
\end{align}
\item 
If $f=0$ and $g\not=0$, then $rank(\mnoi{A}_1)=rank(\mnoi{A})$ and
\begin{align}
\mnoi{A}_1^{\dag}
&=\mnoi{A}^{\dag}
-\mu^{-1}\vnoi{d}
(g\vnoi{d}^*\mnoi{A}^\dag+\bar{\lambda}\vnoi{e}^*
)
+\mu^{-1}\vnoi{g}
(-d\vnoi{e}^*+\lambda\vnoi{d}^*\mnoi{A}^\dag
)
\\
\mnoi{A}_1^\perp&=
\mnoi{A}^\perp-
\frac{
|\lambda|(\sqrt{\mu}-|\lambda|)\vnoi{g}+\lambda g\vnoi{d}
}
{g|\lambda|\sqrt{\mu}
}
\vnoi{g}^*\mnoi{A}^\perp
	\label{eqn:updateperp3}\\
|\mnoi{A}_1^\ddag|&= 
|\mnoi{A}^\ddag|
\frac{(\bar{\lambda}-\lambda)|\lambda|^2 +\lambda \mu}{\mu|\lambda|\sqrt{\mu}}
	\label{eqn:updatedet3}
\,.
\end{align}
\item 
If $f\not=0$ and $g=0$, then $rank(\mnoi{A}_1)=rank(\mnoi{A})$ and
\begin{align}
\mnoi{A}_1^{\dag}
&=\mnoi{A}^{\dag}
-\nu^{-1}(f\mnoi{A}^\dag\vnoi{e}+\bar{\lambda}\vnoi{d}
)\vnoi{e}^*
+\nu^{-1}(-e\vnoi{d}+\lambda\mnoi{A}^\dag\vnoi{e}
)\vnoi{f}^*
\\
\mnoi{A}_1^\perp&=\mnoi{A}^\perp -
\mnoi{A}^\perp\vnoi{f}
\frac{
(|\lambda|(\sqrt{\nu}-|\lambda|)\vnoi{f}+\bar{\lambda} f\vnoi{e})^*}
{f|\lambda|\sqrt{\nu}}
	\label{eqn:updateperp4}\\
|\mnoi{A}_1^\ddag|&=
|\mnoi{A}^\ddag|
\frac{(\lambda-\bar{\lambda})|\lambda|^2 +\bar{\lambda} \nu}{\nu|\lambda|\sqrt{\nu}}
	\label{eqn:updatedet4}
\,.
\end{align}
\item 
If $f\not=0$ and $g\not=0$, then $rank(\mnoi{A}_1)=rank(\mnoi{A})+1$ and
\begin{align}
\mnoi{A}_1^{\dag}
&=\mnoi{A}^{\dag}
-f^{-1}\vnoi{d}\vnoi{f}^*
+g^{-1}\vnoi{g}(-\vnoi{e}^*+\lambda f^{-1}\vnoi{f}^*)\\
\mnoi{A}_1^\perp&=\mnoi{A}^\perp -(1/\sqrt{gf})\vnoi{g}\vnoi{f}^*
	\label{eqn:updateperp5}\\
|\mnoi{A}_1^\ddag|
&=|\mnoi{A}^\ddag|\left[1+(g^{-1}f^{-1}-(1/\sqrt{gf}))
\vnoi{g}^*\mnoi{A}^\perp\vnoi{f}
\right]
	\label{eqn:updatedet5}
\,.
\end{align}
\end{enumerate}
The cost to compute $\mnoi{A}_1^{\dag}$, $\mnoi{A}_1^\perp$, and
$|\mnoi{A}_1^\ddag|$ is $\bigoh(\el^2)$.
					\label{pro:pseudoshermanmorrison}
\end{prop}
\begin{pf} 
The overall method,
update rules for $rank(\mnoi{A}_1)$, and update rules for $\mnoi{A}_1^\dag$
are taken from \cite{BA-BA-TR:2003}, who also list the useful
properties
\begin{equation}
\begin{split}
\begin{array}{ccccc}
\vnoi{c}^*\vnoi{d}=\vnoi{e}^*\vnoi{b}=\lambda-1,
&\vnoi{b}^*\vnoi{f}=f,
&\vnoi{c}^*\vnoi{g}=g,
&\vnoi{d}^*\vnoi{g}=0,
&\vnoi{e}^*\vnoi{f}=0,
\end{array}\\
\begin{array}{cccc}
\mnoi{A}^\dag\mnoi{A}\vnoi{d}=\vnoi{d},
&\mnoi{A}\mnoi{A}^\dag\vnoi{e}=\vnoi{e},
&\mnoi{A}^*\vnoi{f}=\mnoi{A}^\dag\vnoi{f}=0,
&\mnoi{A}\vnoi{g}=(\mnoi{A}^\dag)^*\vnoi{g}=0.
\end{array}
\end{split}
			\label{eqn:pinvobs}
\end{equation}
They give update rules for the row and column spans of $\mnoi{A}_1$,
which we translate into update rules for $\mnoi{A}^\perp$. The cases
(\ref{eqn:updateperp1}), (\ref{eqn:updateperp2}), and
(\ref{eqn:updateperp5}) follow directly.
Corresponding to
(\ref{eqn:updateperp3}), their update rule is that the row span of
$\mnoi{A}^\perp$ should be extended (orthogonally) by $\vnoi{d}$ and
then reduced by projecting orthogonal to $\vnoi{p}$. We translate this
into a (Householder) reflection of the vector $\vnoi{g}$ into a vector
in the span of $\vnoi{d}$ and $\vnoi{g}$ perpendicular to
$\vnoi{p}$. Adjusting these vectors to have equal norm and real inner
product yields the reflection of
the vector $\bar{\lambda}\sqrt{\mu}\,\vnoi{g}$ to
$-|\lambda|(g\vnoi{d}-\bar{\lambda}\vnoi{g})$, resulting in
\begin{equation}
\left(\mnoi{I}-
\frac{2
(\bar{\lambda}\sqrt{\mu}\vnoi{g}+|\lambda|(g\vnoi{d}-\bar{\lambda}\vnoi{g}))
	(\bar{\lambda}\sqrt{\mu}\vnoi{g}+|\lambda|(g\vnoi{d}-\bar{\lambda}\vnoi{g}))^*
}{
\|(\bar{\lambda}\sqrt{\mu}\vnoi{g}+|\lambda|(g\vnoi{d}-\bar{\lambda}\vnoi{g})) \|^2
}
\right)\mnoi{A}^\perp
\,,
\end{equation}
which simplifies to (\ref{eqn:updateperp3}). 
To obtain (\ref{eqn:updateperp4}) we use the same process, extending
the column span by $\vnoi{e}$ and then projecting orthogonal to
$\vnoi{q}$ by a reflection of $\lambda\sqrt{\nu}\,\vnoi{f}$ to
$-|\lambda|(f\vnoi{e}-\lambda\vnoi{f})$. 

To derive the update rules for $|\mnoi{A}_1^\ddag|$, first add the
update rules for $\mnoi{A}_1^\dag$ and $\mnoi{A}_1^\perp$ and then take
the determinant. On the right hand side factor out a copy of
$\mnoi{A}^\ddag$ leaving a low-rank perturbation of the identity, to
which we can apply Proposition~\ref{pro:pertrankdet}. To simplify the
results, we use (\ref{eqn:pinvdefs}), (\ref{eqn:pinvobs}), and the
further observations
\begin{equation}
\begin{array}{ccc}
(\mnoi{A}^\ddag)^{-1}\vnoi{d}=\vnoi{b}-\vnoi{f},
&(\mnoi{A}^\ddag)^{-1}\mnoi{A}^\dag\vnoi{e}=\vnoi{c}-\vnoi{g},
&(\mnoi{A}^\ddag)^{-1}\vnoi{g}=(\mnoi{A}^\perp)^*\vnoi{c},\\
\vnoi{e}^*(\mnoi{A}^\ddag)^{-1}=\vnoi{c}^*-\vnoi{g}^*,
&\vnoi{f}^*(\mnoi{A}^\ddag)^{-1}=\vnoi{b}^*(\mnoi{A}^\perp)^*.
\end{array}
\end{equation}
To obtain (\ref{eqn:updatedet1}) we compute
\begin{multline}
|\mnoi{A}_1^\ddag|=|\mnoi{A}^\ddag|
\left|\mnoi{I}
-d^{-1}\vnoi{b}\vnoi{d}^*\mnoi{A}^\dag
+((1/\sqrt{de})\vnoi{b}-e^{-1}\vnoi{e}
+d^{-1}e^{-1}(\vnoi{d}^*\mnoi{A}^\dag\vnoi{e})\vnoi{b})\vnoi{e}^*
\right|\\
=|\mnoi{A}^\ddag|
\left|\begin{array}{cc}
1-d^{-1}\vnoi{d}^*\mnoi{A}^\dag\vnoi{b}
	&\vnoi{d}^*\mnoi{A}^\dag((1/\sqrt{de})\vnoi{b}-e^{-1}\vnoi{e}
+d^{-1}e^{-1}(\vnoi{d}^*\mnoi{A}^\dag\vnoi{e})\vnoi{b})
	\\
-d^{-1}\vnoi{e}^*\vnoi{b}
	&1+\vnoi{e}^*((1/\sqrt{de})\vnoi{b}-e^{-1}\vnoi{e}
+d^{-1}e^{-1}(\vnoi{d}^*\mnoi{A}^\dag\vnoi{e})\vnoi{b})
      \end{array}
\right|\\
=|\mnoi{A}^\ddag|
\left|\begin{array}{cc}
0
	&\vnoi{d}^*\mnoi{A}^\dag(1/\sqrt{de})\vnoi{b}
	\\
d^{-1}
	&\vnoi{e}^*((1/\sqrt{de})\vnoi{b}
+d^{-1}e^{-1}(\vnoi{d}^*\mnoi{A}^\dag\vnoi{e})\vnoi{b})
      \end{array}
\right|
=|\mnoi{A}^\ddag| (-(1/\sqrt{de}))\,.
\end{multline}
For (\ref{eqn:updatedet2}) we have 
$
|\mnoi{A}_1^\ddag|=|\mnoi{A}^\ddag|
\left|\mnoi{I}-\lambda^{-1}\vnoi{b}\vnoi{e}^*\right|
=|\mnoi{A}^\ddag|(1-\lambda^{-1}\vnoi{e}^*\vnoi{b})
=|\mnoi{A}^\ddag|\lambda^{-1}\,.
$
To obtain (\ref{eqn:updatedet3}) we compute 
\begin{multline}
|\mnoi{A}^\ddag|
\left|\mnoi{I}
+(\mnoi{A}^\ddag)^{-1}\left(
\vnoi{d}(-\mu^{-1}(g\vnoi{d}^*\mnoi{A}^\dag+\bar{\lambda}\vnoi{e}^*)
	-\frac{\lambda\vnoi{g}^*\mnoi{A}^\perp}{|\lambda|\sqrt{\mu}})
\right.\right.\\\left.\left.
+\vnoi{g}(\mu^{-1}(-d\vnoi{e}^*+\lambda\vnoi{d}^*\mnoi{A}^\dag)
	-\frac{(\sqrt{\mu}-|\lambda|)\vnoi{g}^*\mnoi{A}^\perp}{g\sqrt{\mu}})
\right)
\right|\\
=|\mnoi{A}^\ddag|
\left|\begin{array}{cc}
1+(-\mu^{-1}(g\vnoi{d}^*\mnoi{A}^\dag+\bar{\lambda}\vnoi{e}^*))\vnoi{b}
	&(\mu^{-1}(-d\vnoi{e}^*+\lambda\vnoi{d}^*\mnoi{A}^\dag))^*\vnoi{b}	\\
(-\lambda\vnoi{g}^*\mnoi{A}^\perp/|\lambda|\sqrt{\mu})(\mnoi{A}^\perp)^*\vnoi{c}
	&1-((\sqrt{\mu}-|\lambda|)\vnoi{g}^*\mnoi{A}^\perp/g\sqrt{\mu})^*(\mnoi{A}^\perp)^*\vnoi{c}
      \end{array}
\right|\\
=|\mnoi{A}^\ddag|
\left|\begin{array}{cc}
\bar{\lambda}/\mu
	&d/\mu	\\
-\lambda g/|\lambda|\sqrt{\mu}
	&|\lambda|/\sqrt{\mu}
      \end{array}
\right|
=|\mnoi{A}^\ddag|
\frac{(\bar{\lambda}-\lambda)|\lambda|^2 +\lambda \mu }{\mu|\lambda|\sqrt{\mu}}
\,.
\end{multline}
A similar calculation yields (\ref{eqn:updatedet4}).
To obtain (\ref{eqn:updatedet5}) we compute
\begin{multline}
|\mnoi{A}^\ddag|
\left|\mnoi{I}
+(\mnoi{A}^\ddag)^{-1}\left(
-f^{-1}\vnoi{d}\vnoi{f}^*
+
\vnoi{g}(g^{-1}(-\vnoi{e}^*+\lambda f^{-1}\vnoi{f}^*)
-(1/\sqrt{gf})\vnoi{f}^*)
\right)
\right|\\
=|\mnoi{A}^\ddag|
\left|\begin{array}{cc}
1
	&\vnoi{f}^*(\mnoi{A}^\perp)^*\vnoi{c}
\\
g^{-1}f^{-1}(\lambda-1)
	&1+(g^{-1}\lambda f^{-1}-(1/\sqrt{gf}))
		\vnoi{f}^*(\mnoi{A}^\perp)^*\vnoi{c}
      \end{array}
\right|\\
=|\mnoi{A}^\ddag|
(1+((-(1/\sqrt{gf}))
+g^{-1}f^{-1})\vnoi{f}^*(\mnoi{A}^\perp)^*\vnoi{c}
)\\
=|\mnoi{A}^\ddag|\left[1+(g^{-1}f^{-1}-(1/\sqrt{gf}))
\vnoi{g}^*\mnoi{A}^\perp\vnoi{f}
\right].
\end{multline}
\qed\end{pf}

When $\mnoi{A}$ and $\mnoi{A}_1$ are nonsingular,
(\ref{eqn:likeshermanmorrison}) is the Sherman-Morrisson Formula (see e.g.\
\cite{GOL-LOA:1996}). For our application we need the singular vectors in $\mnoi{A}^\perp$,
rather than $\mnoi{A}^\perp$ itself, but then only when
$rank(\mnoi{A}^\perp)\le 3$. These singular vectors can be extracted
by a simple modification of the power method with deflation.

\raggedright
\small


\end{document}